\documentclass[sn-mathphys,Numbered]{sn-jnl}% Math and Physical Sciences Reference Style
\usepackage{graphicx}%
\usepackage{multirow}%
\usepackage{amsmath,amssymb,amsfonts}%
\usepackage{amsthm}%
\usepackage{mathrsfs}%
\usepackage[title]{appendix}%
\usepackage{xcolor}%
\usepackage{textcomp}%
\usepackage{manyfoot}%
\usepackage{booktabs}%
\usepackage{algorithm}%
\usepackage{algorithmicx}%
\usepackage{algpseudocode}%
\usepackage{listings}%
\usepackage{chemformula}
\usepackage{here}
\usepackage{color}
\usepackage{soul}
\usepackage{threeparttable}

\theoremstyle{thmstyleone}%
\theoremstyle{thmstyletwo}%

\theoremstyle{thmstylethree}%

\begin{document}

\title[Article Title]{Designing Superionic Conductors Using Tetrahedrally Packed Structures}


\author*[1]{\fnm{Tomoyasu} \sur{Yokoyama}}\email{yokoyama.tomoyasu@jp.panasonic.com}
\author[1]{\fnm{Kazuhide} \sur{Ichikawa}}
\author[1]{\fnm{Takuya} \sur{Naruse}}
\author[1]{\fnm{Kosei} \sur{Ohura}}
\author[1]{\fnm{Yukihiro} \sur{Kaneko}}

\affil*[1]{Technology Division, Panasonic Holdings Corporation,  1006 Kadoma, Kadoma City, Osaka 571-8508, Japan}


\abstract{
In the pursuit of advanced energy storage solutions, the crystal structure of ionic conductors plays a pivotal role in facilitating ion transport. 
The conventional structural design principle that compounds with the body-centered cubic (BCC) anionic frameworks have high ionic conductivity is well known. 
We have extended the conventional design principle by uncovering that many of the anionic frameworks of Ag-ion conductors are characterized by tetrahedrally packed (TP) structures.
Leveraging our findings, we have virtually screened TP framework compounds, uncovering their intrinsic potential for superior ionic conductivity through first-principles molecular dynamics simulations.
Our design principle is applicable to Ag$^+$ and other mobile ions, including Li$^+$ and F$^-$. 
We proposed the Met2Ion method to generate ionic crystal structures using metal crystal structures as templates and demonstrated that new ionic conductors with TP frameworks can be discovered.
This work paves the way for the discovery and development of next-generation energy storage materials with enhanced performance.
}




\maketitle

\section{Introduction}\label{sec1}
The incessant drive for innovation in energy storage technologies has positioned ionic conductivity as a cornerstone material property, especially for applications in fuel cells and batteries \citep{Knauth2009,Minh1993}. 
In this regard, the structural design of ion conductors has emerged as a critical area of research, with the aim of enhancing the performance and safety of these devices.
The significance of the crystal structure in influencing ionic transport is well-recognized, and a substantial body of work has been devoted to the exploration of Li-ion conductors, given their central role in all-solid-state Li-ion batteries \citep{Deiseroth2008,Kato2016,Asano2018,Li2023,Tanaka2023}.

\begin{figure}[t]%
\centering
\includegraphics[width=0.9\textwidth]{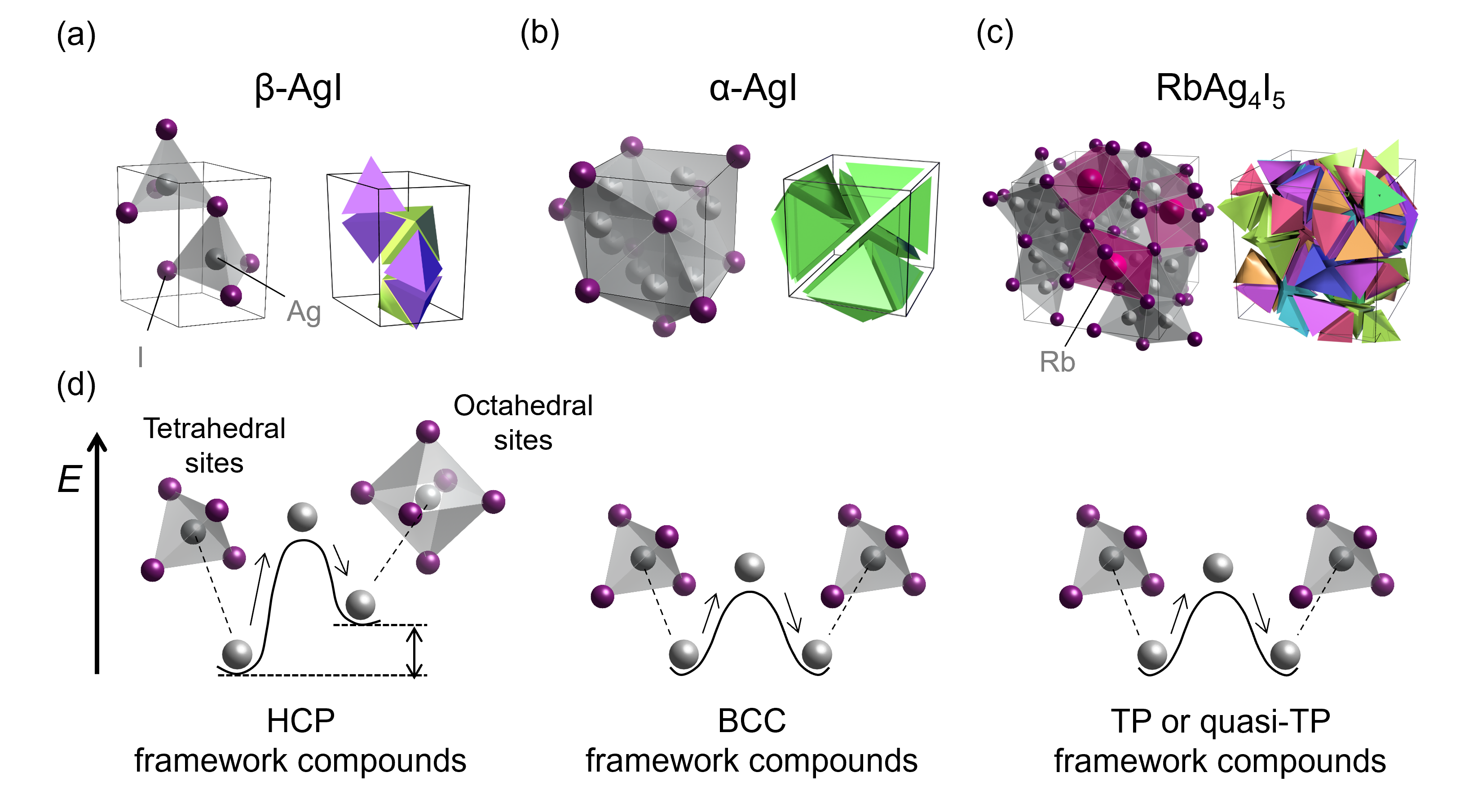}
\caption{Crystal structures and polyhedral tiling of (a) $\beta$-AgI with a hexagonal close-packed (HCP) framework, (b) $\alpha$-AgI with a body-centered cubic (BCC) framework, and (c) \ch{RbAg4I5} with a quasi-tetrahedrally packed (TP) framework. (d) Conceptual diagram illustrating the migration barriers for Ag-ion diffusion within these frameworks.}\label{fig1}
\end{figure}

Previous research has established that the topology of the framework composed of non-mobile ions in ionic conductors affects its ability to mobile ions \citep{Wang2015}. 
Specifically, studies have highlighted that body-centered cubic (BCC) frameworks, composed solely of tetrahedral sites, enable more efficient ion transport than other structures like face-centered cubic (FCC) or hexagonal close-packed (HCP) frameworks. 
As shown in Figure \ref{fig1}a, the tetrahedron in the HCP structure shares a face with one tetrahedron and three octahedra. 
The mobile ions at the tetrahedral site must pass through an octahedral site to diffuse outward.
Different site energies, including those of the tetrahedral and octahedral sites, lead to greater energy gaps to the transition state for ion diffusion, which leads to lower ionic conductivity. 
In contrast, the BCC structure is composed of tetrahedra only, and mobile ions can diffuse from one tetrahedral site to another tetrahedral site without passing through octahedral sites, as shown in Figure \ref{fig1}b. 
Therefore, because the site energy remains constant, the energy barrier for ionic diffusion is lower in the BCC anionic framework than that in the HCP framework. 
In fact, the experimental values for AgI ionic conductivity within the HCP and BCC anionic frameworks are 10$^{-4}$ S/cm and 1 S/cm, respectively, at about 420 K \citep{Tubandt1914}, showing significantly higher Ag-ion conductivity with a BCC framework.
This principle has also been applied to Li-ion conductors like \ch{Li10GeP2S12}, which possess BCC frameworks and offer high Li-ion conductivity, promising safer solid electrolyte candidates for all-solid-state Li-ion batteries \citep{Kato2016,Li2023}. 
These insights are crucial for designing materials with high ionic conductivity for advanced energy storage applications.

The following questions then arise: What is the ideal structure for high ionic conductivity, and could it be a BCC framework structure?
To explore this, our research turns to the Ag-ion conductors, known for their exceptional ionic conductivities, with the aim of identifying framework structures that could serve as templates for new superionic materials. 
While research in the literature so far has extensively examined the topology of Li-ion conductors, Ag-ion conductors have received less attention. 
Yet, compounds like \ch{RbAg4I5} exhibit remarkably high Ag-ion conductivity \citep{Owens1967}, about ten times greater than their Li-ion conductors \citep{Kato2016,Li2023}. 
By delving into the structural intricacies of Ag-ion conductors, we aim to broaden the understanding of ionic conductivity and potentially discover universal design principles applicable to various ionic conductors.

Our study introduces a new structural design principle highlighting the significance of tetrahedrally packed (TP) frameworks for high ionic conductivity. 
Our analysis revealed that many of the anionic frameworks of high Ag-ion conductors are characterized by TP structures.
The TP structure, which is composed entirely of tetrahedral tiling, is regarded as a superordinate concept to the BCC structure.
For example, the \ch{RbAg4I5} structure shown in Figure \ref{fig1}c features an anionic framework of the $\beta$-\ch{Mn}-type structure, which belongs to the TP structure family.
While it was previously known in a limited context that argyrodite-type ionic conductors exhibit high conductivity due to their tetrahedrally close-packed (TCP) framework—a specialized subgroup of TP structures—there hasn't been much discussion about other TP frameworks \citep{Kuhs1979,Morgan2021,Han2024}.
Armed with the insight that TP framework structures could be key to achieving high ionic conductivity, we embarked on a systematic screening of TP framework compounds from experimental databases.
Our first-principles molecular dynamics (FPMD) calculations confirmed that these compounds exhibit high ionic conductivity across various mobile ion species, not just limited to Ag$^{+}$ ions.
This finding is significant as it suggests that the TP framework is a universal design motif for high conductivity, applicable to both anion and cation conductors.

To further the discovery of unknown superionic conductors with this new design principle, we introduced the Met2Ion method, a novel crystal structure design approach inspired by the structural parallels between metallic and ionic crystals.
Our Met2Ion method transcends the capabilities of conventional template-based prediction methods, enabling the prediction of previously unknown ionic crystal structures.
Employing the Met2Ion method, we successfully identified novel Li-ion compounds with TP frameworks that are exhibit high ionic conductivity, demonstrating our method's efficacy in virtual screening. 
The implications of this work are profound, as it not only enhances the understanding of ionic conductivity but also provides a powerful tool for the discovery of novel superionic conductors. 
Our study thus sets a new precedent for the efficient development of ionic conductors, paving the way for the next generation of energy storage technologies.

\section{Results}\label{sec2}
\subsection{Identification of framework structures of Ag-ion conductors}\label{subsec2}

\begin{figure}[H]
\centering
\includegraphics[width=0.9\textwidth]{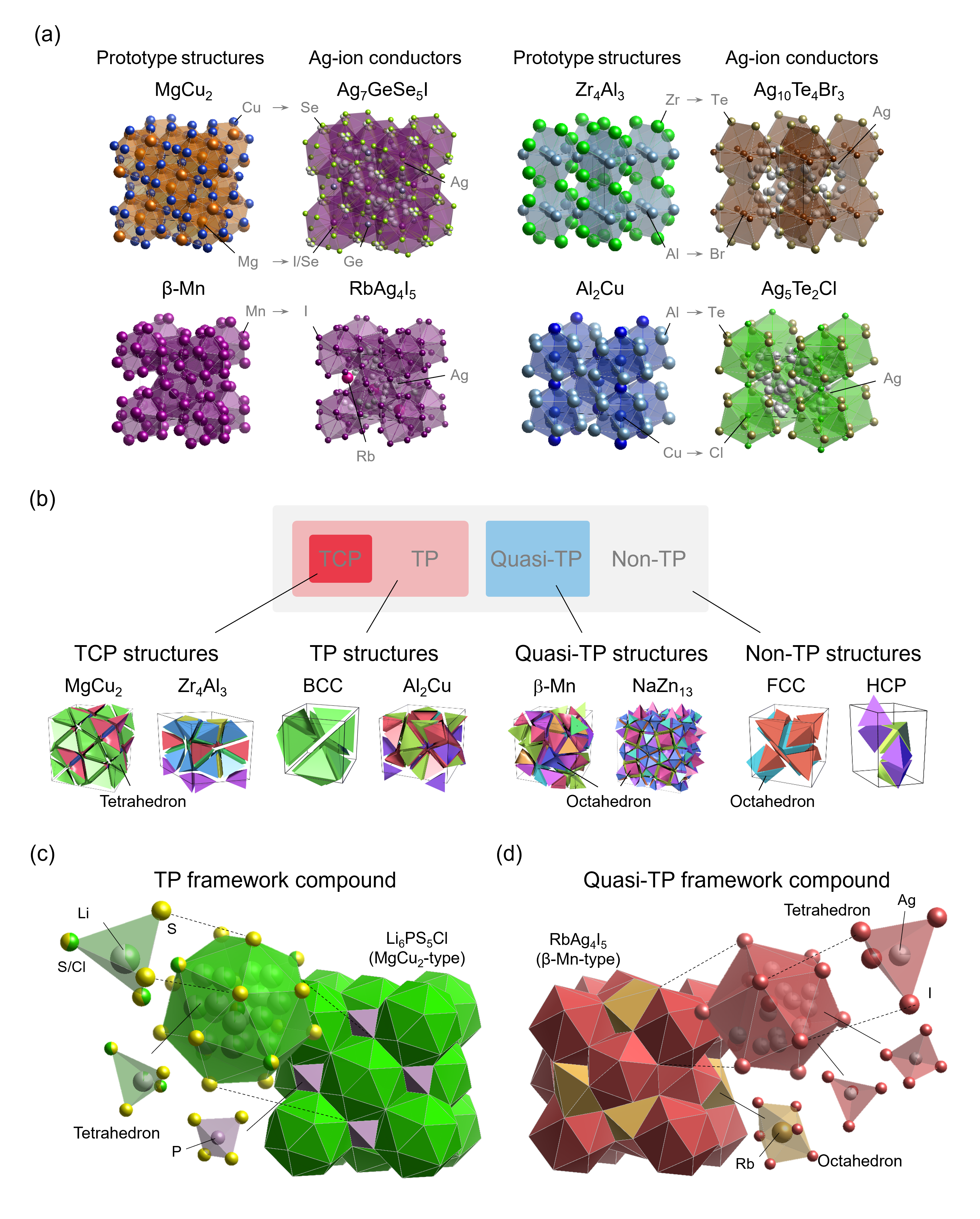}
\caption{(a) Prototype structures and their corresponding Ag-ion conductors, including \ch{Ag7GeSe5I} with a \ch{MgCu2}-type framework, \ch{Ag10Te4Br3} with a \ch{Zr4Al3}-type framework,  \ch{RbAg4I5} with a $\beta$-Mn-type framework, and \ch{Ag5Te2Cl} with an \ch{Al2Cu}-type framework. (b) Classification of structures based on space-filling polyhedra and example of polyhedral tiling of tetrahedrally close-packed (TCP), tetrahedrally packed (TP), quasi-TP, and non-TP structures. (c) A crystal structure composed of \ch{Li6PS5Cl} with a \ch{MgCu2}-type framework. (d) A crystal structure of \ch{RbAg4I5} with a $\beta$-Mn-type framework.}\label{fig2}
\end{figure}

Our research began by examining the structure of Ag-ion conductors, which are known for their outstanding ionic conductivity. 
We sourced the crystal structures of these conductors from the Inorganic Crystal Structure Database (ICSD) \citep{Bergerhoff1983} and analyzed them to identify their anionic framework prototypes. 
Through this analysis, we found that many of these conductors have BCC anionic frameworks, known to yield high ionic conductivity. 
For instance, AgI and \ch{Ag3SI} both have BCC anionic frameworks and demonstrate high Ag-ion conductivities of 1 S/cm at 420 K \citep{Tubandt1914} and 3.8$\times$10$^{-2}$ S/cm \citep{Gombotz2021}, respectively. 
These insights, compiled in Table S1, reconfirm the promising role of BCC anionic frameworks in crafting materials with superior conductivity, aligning with established design principles.

In addition, our research identified superionic conductors that are different from the BCC framework. 
For instance, \ch{Ag7GeSe5I} exhibits an Ag-ion conductivity of 4.3$\times10^{-2}$ S/cm \citep{Belin2001} and has an anionic framework of \ch{MgCu2}-type structure called cubic Laves phase.
\ch{Ag10Te4Br3} shows an Ag-ion conductivity of 1.4$\times$10$^{-2}$ S/cm at room temperature \citep{Lange2006} and has a \ch{Zr4Al3}-type framework. 
\ch{RbAg4I5}, with a conductivity of 2.1$\times$10$^{-1}$ S/cm at room temperature \citep{Owens1967} and has a $\beta$-Mn-type framework. 
Additionally, \ch{Ag5Te2Cl} has an Ag-ion conductivity of 1.0$\times$10$^{-1}$ S/cm at 341 K \citep{Nilges2004} and has a \ch{Al2Cu}-type framework. 
These diverse structures, illustrated in Figure \ref{fig2}a, all represent significant departures from the BCC framework, demonstrating the variety of crystal structures leading to high ionic conductivity.

\begin{table}[htbp]
\caption{List of reported tetrahedrally packed (TP), tetrahedrally close-packed (TCP), and quasi-TP structures. In the ``Category" column, ``TP", ``TCP", and ``q-TP" refer to the TP, TCP, and quasi-TP structure, respectively. ${N_{atom}}$ means the number of atoms in the unit cell. ``Relevant ionic crystal" refers to a representative ionic crystal that has the same anionic or cationic framework as the structure of its metallic crystal. Elements that compose the framework in the compound are shown in bold.}\label{tab1}%
\begin{tabular}{@{}llllll@{}}
\hline
Category &Structure type & Space group & ${N_{atom}}$  & Relevant ionic crystal & Ref.\\
\hline
TP &\ch{BCC} &$Im\overline{3}m$ &2 & \ch{Ag\textbf{I}}, \ch{\textbf{CaTi}O3}& \citep{Bonneau2015}\\
TCP &\ch{MgCu2} &$Fd\overline{3}m$ &24 & \ch{Ag7Ge\textbf{Se5I}}, \ch{\textbf{MgAl2}O4}& \citep{Sikiric2010,Bonneau2015}\\
TCP &\ch{MgZn2} &$P6_{3}/mmc$ &12 & \ch{Li6Si\textbf{O4Cl2}}, \ch{\textbf{NaBe4Sb}O7}& \citep{Sikiric2010,Bonneau2015}\\
TCP &\ch{MgNi2} &$P6_{3}/mmc$ &24 & \ch{\textbf{Mg3BeAl8}O16}& \citep{Sikiric2010}\\
TCP &\ch{Cr3Si} &$Pm\overline{3}m$ &8 & \ch{Cu6\textbf{Te3S}}, \ch{\textbf{Y3Al5}O12}& \citep{Sikiric2010,Bonneau2015}\\
TCP &\ch{Zr4Al3} &$P6/mmm$ &7 & \ch{Ag10\textbf{Te4Br3}}, \ch{\textbf{BaTi2Fe4}O11}& \citep{Sikiric2010,Bonneau2015}\\
TCP &\ch{Mg4Zn7} &$C2/m$ &110 & -& \citep{Sikiric2010}\\
TCP &\ch{Co8Mn9Si3} &$Pnnm$ &74 & -& \citep{Sikiric2010}\\
TCP &\ch{Na52Au81Si29} &$Im\overline{3}$ &162 & -& \citep{Sikiric2010}\\
TCP &\ch{Ti2(Ni,Al,Ti)3} &$C2/m$ &50 & -& \citep{Sikiric2010}\\
TCP &\ch{K7Cs6} &$P6_{3}/mmc$ &26 & \ch{\textbf{SrFe12}O19}& \citep{Sikiric2010}\\
TCP &\ch{Th6Cd7} &$Pbam$ &26 & -& \citep{Sikiric2010}\\
TCP &\ch{W6Fe7} &$R\overline{3}m$ &39 & \ch{Ag19\textbf{Te6Br7}}, \textbf{K$_{2}$Al$_{11-x}$}O$_{17}$& \citep{Sikiric2010}\\
TCP &\ch{Nb48Ni39Al13} &$Pnma$ &52 & Ag$_{19}$\textbf{Br$_{5.438}$I$_{1.562}$Te$_{6}$}& \citep{Sikiric2010}\\
TCP &\ch{Mg23Al30} &$R\overline{3}$ &159 & -& \citep{Sikiric2010}\\
TCP &\ch{Mn77Fe4Si19} &$C2$ &110 & -& \citep{Sikiric2010}\\
TCP &\ch{Cr9Mo21Ni20} &$Pnma$ &56 & -& \citep{Sikiric2010}\\
TCP &\ch{MoNi} &$P2_{1}2_{1}2_{1}$ &56 & -& \citep{Sikiric2010}\\
TCP &\ch{Mn82Si18} &$Immm$ &186 & -& \citep{Sikiric2010}\\
TCP &$\beta$-\ch{U} &$P4_{2}/mnm$ &30 & \ch{\textbf{N}F3}& \citep{Sikiric2010}\\
TP &\ch{Al2Cu} &$I4/mcm$ &12 & \ch{Ag5\textbf{Te2Cl}}, \ch{\textbf{Pb3}O4}& \citep{Bonneau2015}\\
TP &\ch{CaCu5} &$P6/mmm$ &6 & \ch{\textbf{Cd3}PCl3}& \citep{Bonneau2015}\\
TP &\ch{AlB2} &$P6/mmm$ &3 & \ch{\textbf{PbSb2}O6}& \citep{Bonneau2015}\\
TP &\ch{ThSi2} &$I4_{1}/amd$  &12 & \ch{\textbf{Na3}N8}& \citep{Bonneau2015}\\
TP &\ch{FeB} &$Pnma$ &8 & \ch{Li2\textbf{IO}H}, \ch{\textbf{BaS}O4}& \citep{Bonneau2015}\\
TP &\ch{CrB} &$Cmcm$ &8 & \ch{Li(\textbf{N}H3)\textbf{I}}, \ch{\textbf{KCl}O3}& \citep{Bonneau2015}\\
TP &\ch{CeNi3} &$P6_{3}/mmc$ &24 & -& \citep{Bonneau2015}\\
TP &\ch{PuNi3} &$R\overline{3}m$ &36 & \ch{\textbf{CuTi2}S4}& \citep{Bonneau2015}\\
TP &\ch{Ce2Ni7} &$P6_{3}/mmc$ &36 & -& \citep{Bonneau2015}\\
TP &\ch{Th2Ni17} &$P6_{3}/mmc$ &38 & -& \citep{Bonneau2015}\\
TP &\ch{Th2Zn17} &$R\overline{3}m$ &57 & -& \citep{Bonneau2015}\\
q-TP &$\beta$-\ch{Mn} &$P4_{1}32$ &20 & \ch{RbAg4\textbf{I5}}, \ch{\textbf{Ag4}Te\textbf{S}O4}& \citep{Karlsen1992}\\
q-TP &\ch{BaCd11} &$I4_{1}/amd$ &48 & -& \citep{Bonneau2015}\\
q-TP &\ch{NaZn13} &$Fm\overline{3}c$ &112 & \ch{Mg3B7\textbf{O13Cl}}, \ch{\textbf{Ag13Os}O6}& \citep{Bonneau2015}\\
q-TP &\ch{ThMn12} &$I4/mmm$ &26 & -& \citep{Bonneau2015}\\
\hline
\end{tabular}
\end{table}

To comprehend the characteristics of these prototype structures that diverge from the BCC framework, we encountered insights from Bonneau \textit{et al.}'s study on metallic crystals \citep{Bonneau2015}. 
Their study indicated that structures such as \ch{MgCu2}, \ch{Zr4Al3}, and \ch{Al2Cu} are classified as tetrahedrally packed (TP) structures, which can be filled with slightly distorted tetrahedra \citep{OKeeffe1998}. 
The BCC structure is a unique TP structure composed of all congruent tetrahedra.
Meanwhile, other structures, such as the \ch{MgCu2}-type, are also categorized within the TP family but are specifically known as Frank-Kasper structures or tetrahedrally close-packed (TCP) structures—a concept first introduced by Frank and Kasper in the 1950s.
Specifically, a structure is recognized as a TCP structure if it features a pattern where five or six tetrahedra meet at an edge \citep{OKeeffe1998,Shoemaker1986}. 
These TCP structures have unique coordination polyhedra, referred to as Frank-Kasper polyhedra.
Their report by Bonneau \textit{et al.} expands on TP structures, including quasi-TP structures that are predominantly tetrahedral but may contain other polyhedra, such as in $\beta$-Mn and \ch{NaZn13}. 

Figures \ref{fig2}b display various structural types, including TCP, TP, quasi-TP, and non-TP structures.
TP structures like BCC and \ch{MgCu2}-type are filled with tetrahedra, unlike FCC and HCP structures. 
The \ch{Li6PS5Cl} structure has a \ch{MgCu2}-type anionic framework with multiple tetrahedra \citep{Morgan2021}, as seen in Figure \ref{fig2}c. 
Here, Li atoms are located at two tetrahedral sites within a Frank-Kasper polyhedron. 
The tetrahedra around Li atoms form a three-dimensional network without interference from the P atom tetrahedra.
The \ch{RbAg4I5} anionic framework, depicted in Figure \ref{fig2}d, is a quasi-TP structure with six types of tetrahedra and an octahedron, making up 96\% tetrahedral sites. 
The tetrahedra around Ag atoms form a three-dimensional network without interference from the Rb atom octahedra.
Table \ref{tab1} presents a selection of other TP prototype structures and related ionic crystals. 

While the individual anionic frameworks of \ch{Li6PS5Cl} and \ch{RbAg4I5}, as well as their high ionic conductivity, were known \citep{Hull2002,Morgan2021}, our discovery lies in conceptualizing these and other frameworks as the broader TP structure family.
Our findings highlight TP structures as a key to designing
high-performance ionic conductors, extending beyond the BCC structures, due to their potential for lower ion migration barriers.
Therefore, in response to the question presented in the introduction regarding what constitutes an ideal structure for high ionic conductors, TP framework structures have emerged as a promising answer.

\subsection{Screening of TP framework compounds with mobile cations and anions}\label{subsec2}
\begin{figure}[t]%
\centering
\includegraphics[width=1\textwidth]{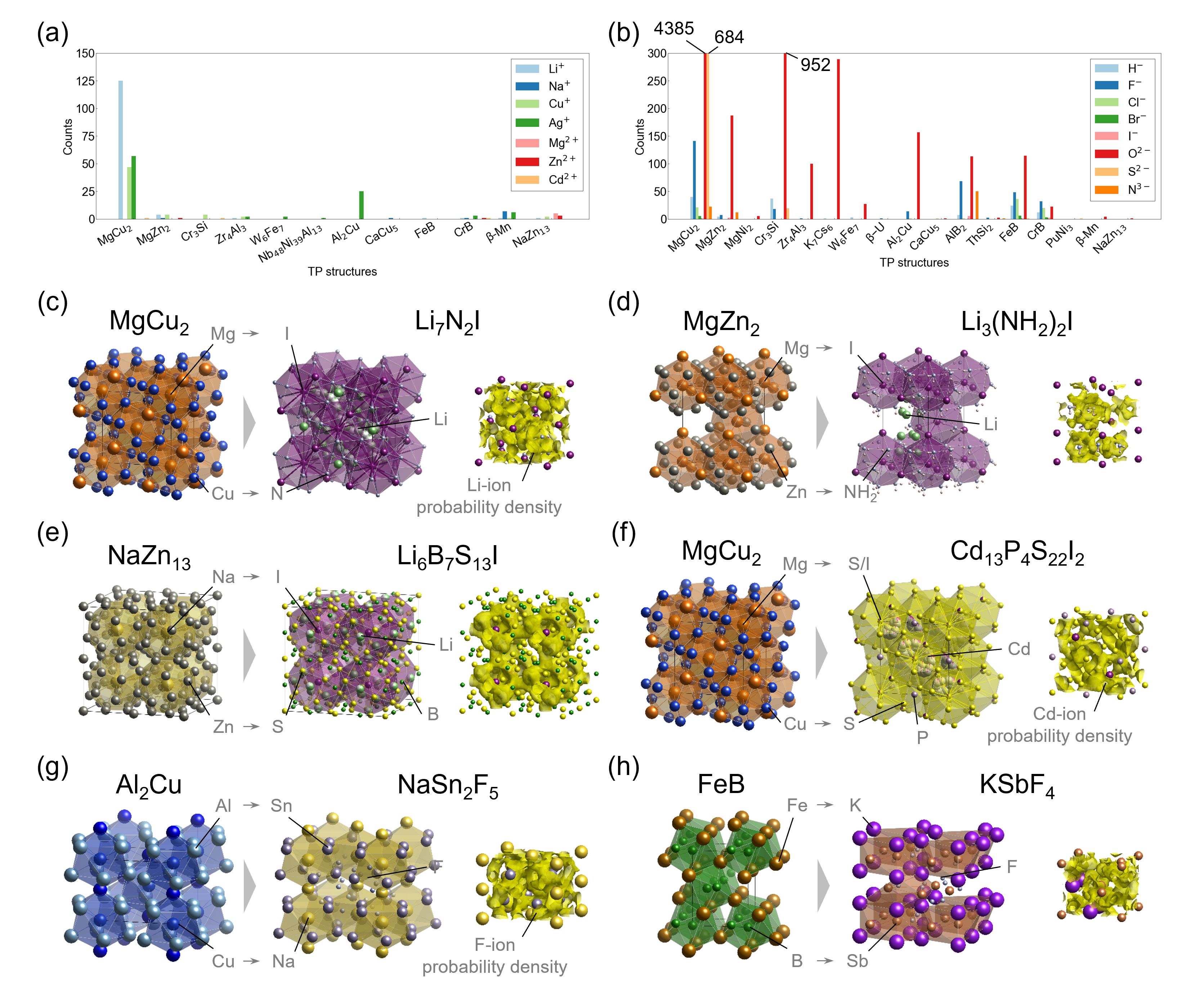}
    \caption{Histograms of (a) tetrahedrally packed (TP) anionic framework compounds and (b) TP cationic framework compounds in the ICSD. Structures of the  TP metallic compounds and TP framework compounds and ionic conductive pathways of TP framework compounds in (c) \ch{Li7N2I} with \ch{MgCu2}-type framework, (d) \ch{Li3(NH2)2I} with \ch{MgZn2}-type framework, (e) \ch{Li6B7S13I} with \ch{NaZn13}-type framework, (f) \ch{Cd13P4S22I2} with \ch{MgCu2}-type framework, (g) \ch{NaSn2F5} with \ch{Al2Cu}-type framework, and (h) \ch{KSbF4} with \ch{FeB}-type framework. }\label{fig3}
\end{figure}

To verify the hypothesis that TP frameworks lead to high ionic conductivity, we analyzed TP compounds from the ICSD using FPMD calculations.
First, we focused on cases where the anion forms the framework and the cation is mobile. 
We extracted compounds with an anionic framework matching the TP structures from the ICSD and evaluated their ionic conductivities through FPMD simulations.
Figure \ref{fig3}a displays a histogram of these compounds, categorized by mobile ion species.
The most commonly identified anionic framework was the \ch{MgCu2}-type structure, with the argyrodite-type ionic structure being a representative example of an iconic compound with this anionic framework.
Argyrodite-type ionic compounds have been recognized for their superionic conductivity in Cu and Ag compounds since the 1970s \citep{Kuhs1978} and similar high Li-ion conductivity was reported in 2008 \citep{Deiseroth2008}, sparking extensive research and a surge in these compounds.

We evaluated the ionic conductivities of TP framework compounds through FPMD calculations.
Table \ref{tab2} summarizes compounds with conductivities over 1.0$\times$10$^{-4}$ S/cm, alongside experimental values.
Complete data is available in Table S2.
Figures \ref{fig3}c-f show the crystal structures of typical TP anionic framework compounds and the mobile ion probability densities from FPMD simulations.
Arrhenius plots from these simulations are in Figure S1.

Argyrodite-type compounds with \ch{MgCu2}-type frameworks, such as \ch{Ag7GeSe5I} and \ch{Li6PS5Cl}, exhibit high theoretical and experimental ionic conductivities. 
\ch{Li7N2I}, another \ch{MgCu2}-type framework compound, which is not argyrodite-type, demonstrates high theoretical conductivity.
Additionally, compounds with TCP structures, including \ch{MgZn2}, \ch{Cr3Si}, \ch{Zr4Al3}, and \ch{W6Fe7} frameworks, show potential for high ionic conductivity.
For instance, \ch{Ag19Te6Br7} features a \ch{W6Fe7}-type framework with Te and Br substituting for W and Fe.
\ch{Li3(NH2)2I} features a \ch{MgZn2}-type framework with I and \ch{NH2} atoms substituting for Mg and Zn, respectively.
Similarly, \ch{Li6SiO4Cl} possesses a \ch{MgZn2}-type framework where Cl and O ions replace Mg and Zn atoms.
Compounds with frameworks composed of TP structures, such as \ch{Ag5Te2Cl} with an \ch{Al2Cu}-type framework, also displayed high ionic conductivities.
\ch{Li6B7S13I}, derived from \ch{NaZn13} with S and I substitutions, showed a theoretical Li-ion conductivity of 2.1$\times$10$^{-3}$ S/cm.
Cd-ion compounds typically exhibit low ionic conductivity due to the large ionic radius and strong Coulomb interactions of Cd$^{2+}$ ions.
However, \ch{Cd13P4S22I2} with a TP framework is theoretically predicted to have high ionic conductivity.
This suggests that TP frameworks can enhance ionic conductivity in compounds with various cations.

\begin{sidewaystable}
\caption{Summary of the calculated and experimental ionic conductivities and activation energies of the tetrahedrally packed (TP) framework compounds. ``Ion" refers to a possible mobile ion in the compound. ``Formula" refers to the compositional formula of the calculated structure. ``ICSD ID" refers to the database collection code. ``Framework type" refers to the structural type of the anionic framework for each compound. $\sigma_{300K}$ and $E_{a,300K}$ mean the ionic conductivity at room temperature and the activation energy as extrapolated from Arrhenius plots based on first-principle molecular dynamics (FPMD) calculations, respectively. $\sigma_{exp.}$ is the reported ionic conductivity. When the temperature is not stated, the ionic conductivity is at room temperature. }\label{tab2}
\begin{tabular*}{\textheight}{@{}llllllll@{}}
\hline
Ion & ICSD ID & Formula & Framework type & $\sigma_{300K}$ (S/cm) & $E_{a,300K}$ (eV) & $\sigma_{exp.}$ (S/cm) & Ref.\\
\hline
{Ag}$^{+}$ & 241154\tnote{a} & \ch{Ag7GeSe5I} & \ch{MgCu2} & 4.6$\times10^{-2}$ & 0.158 & 4.3$\times10^{-2}$  & \citep{Belin2001}\\
{Ag}$^{+}$ & 418781 & \ch{Ag10Te4Br3} & \ch{Zr4Al3} & 3.8$\times10^{-2}$ & 0.169 & 1.4$\times10^{-2}$  &\citep{Lange2006}\\
{Ag}$^{+}$ & 421400 & \ch{Ag5Te2Cl} & \ch{Al2Cu} & 1.5$\times10^{-2}$ & 0.206 & 1.0$\times10^{-1}$ $@$341K & \citep{Nilges2004}\\
{Ag}$^{+}$ & 419160 & \ch{Ag19Te6Br7} & \ch{W6Fe7} & 7.2$\times10^{-2}$ & 0.158 & 1.1$\times10^{-2}$ $@$323K & \citep{Nilges2008}\\
{Ag}$^{+}$ & 27203 & \ch{RbAg4I5} & $\beta$-Mn & 4.8$\times10^{-2}$ & 0.139 & 2.1$\times10^{-1}$  &  \citep{Owens1967}\\
{Cu}$^{+}$ & 418657\tnote{a} & \ch{Cu6PS5Cl} & \ch{MgCu2} & 9.6$\times10^{-4}$ & 0.275 & 4.7$\times10^{-4}$  &  \citep{Kuhs1979}\\
{Cu}$^{+}$ & 88660 & \ch{Cu8GeSe6} & \ch{MgZn2} & 4.6$\times10^{-4}$ & 0.403 & - & - \\
{Cu}$^{+}$ & 427561 & \ch{Cu6Te3S} & \ch{Cr3Si} & 8.5$\times10^{-3}$ & 0.246 & - & - \\
{Cu}$^{+}$ & 610355 & \ch{Cu6SbAs} & \ch{Cr3Si} & 7.7$\times10^{-4}$ & 0.233 & - & - \\
{Cu}$^{+}$ & 140304 & \ch{Cu9Te4Cl3} & \ch{Zr4Al3} & 4.1$\times10^{-2}$ & 0.177 & - & - \\
{Cu}$^{+}$ & 51911 & \ch{KCu4I5} & $\beta$-Mn & 3.2$\times10^{-2}$ & 0.164 & 6.1$\times10^{-1}$ $@$540K &  \citep{Hull2002}\\
{Li}$^{+}$ & 418490\tnote{a} & \ch{Li6PS5Cl} & \ch{MgCu2} & 1.1$\times10^{-2}$ & 0.180 & 5.0$\times10^{-3}$  &  \citep{Yu2018}\\
{Li}$^{+}$ & 85713 & \ch{Li7N2I} & \ch{MgCu2} & 1.1$\times10^{-3}$ & 0.281 & - & - \\
{Li}$^{+}$ & 167528 & \ch{Li3(NH2)2I} & \ch{MgZn2} & 5.3$\times10^{-3}$ & 0.217 & 1.7$\times10^{-5}$  &  \citep{Matsuo2010}\\
{Li}$^{+}$ & 143283 & \ch{Li6SiO4Cl2} & \ch{MgZn2} & 3.5$\times10^{-4}$ & 0.199 & 6.2$\times10^{-6}$ $@$575K  &  \citep{Morscher2021}\\
{Li}$^{+}$ & 143928 & \ch{Li6B7S13I} & \ch{NaZn13} & 2.1$\times10^{-3}$ & 0.247 & 5.0$\times10^{-4}$  &  \citep{Kaup2021}\\
{Cd}$^{2+}$ & 622 & \ch{Cd13P4S22I2} & \ch{MgCu2} & 6.6$\times10^{-4}$ & 0.279 & - & - \\
{F}$^{-}$ & 14136 & \ch{NaSn2F5} & \ch{Al2Cu} & 1.8$\times10^{-3}$ & 0.432 & 4.1$\times10^{-7}$ $@$373K  & \citep{Patro2009}\\
{F}$^{-}$ & 431489 & \ch{KSbF4} & \ch{FeB} & 1.5$\times10^{-3}$ & 0.332 & 1.0$\times10^{-4}$  &  \citep{kawahara2021}\\
\hline
\end{tabular*}
\begin{tablenotes}
\item[a] Since numerous argyrodite-type ionic compounds were identified, Table \ref{tab2} includes only one result per mobile ion species.
\end{tablenotes}
\end{sidewaystable}

We also studied cases where the cation forms the framework and the anion is mobile. 
Figure \ref{fig3}b displays a histogram categorizing these compounds, highlighting a greater prevalence of cationic frameworks compared to anionic frameworks.
Since these compounds with TP cationic framework are typical ionic crystal structures such as spinel-type, pyrochlore-type, and garnet-type, the number of candidate structures is larger than in the case of the anionic framework.
Among the compounds extracted in this case, \ch{NaSn2F5} and \ch{KSbF4}, which have been reported as F-ion conductors, are shown in Figure \ref{fig3}g and \ref{fig3}h as an example. 
\ch{KSbF4} has a distorted \ch{FeB}-type framework in which the Fe and B atoms in \ch{FeB} are replaced with K and Sb atoms, respectively.
\ch{NaSn2F5} has a \ch{Al2Cu}-type framework in which the Al and Cu atoms in \ch{Al2Cu} are replaced with Sn and Na atoms, respectively.
The F-ion conductivities of these compounds at room temperature obtained from the FPMD calculations were over 10$^{-3}$ S/cm.
Due to the large number of candidate structures and the high computational cost, comprehensive screening of anionic conductors with TP framework structures is reserved for our future work.

Additionally, it should be noted that compositional design could enhance the ionic conductivity of TP framework compounds.
For instance, despite having a \ch{MgCu2}-type anionic framework, \ch{Li8N2Te} exhibited negligible Li-ion conduction in our calculations, which can be attributed to a high number of mobile ions occupying tetrahedral sites, leaving few vacancies for ion diffusion. 
FPMD calculations indicate that decreasing the number of Li-ions can lead to enhanced conductivity, as demonstrated in Figure S2.
These findings indicate that TP framework compounds, previously considered low ionic conductors, may possess high conductivity potential with the right compositional design.

We comment here on the effectiveness of FPMD calculations in predicting ionic conductivity.
Although there are some differences between experimental and theoretical ionic conductivities, our computational results align with the overall trends observed in experimental data, as also advocated in the literature \citep{Zhu2018}.
It is true that potential deviations can stem from impurities, grain boundaries, and synthesis-related variations \citep{Nolan2018}; however, FPMD calculations remain a useful predictive tool for ionic conductivities.

Thus, our investigation reveals that TP framework compounds demonstrate high theoretical ionic conductivity, irrespective of the mobile ion being an anion or cation.
These insights offer a novel design principle for ionic conductors, potentially superseding traditional BCC framework guidelines.

\subsection{Search for unknown TP framework compounds via rule-based structure generation}\label{subsec2}
To bridge the gap in discovering unsynthesized TP framework compounds with high ionic conductivity, we developed the Met2Ion method for designing such materials.
In the preceding section, we identified high ionic conductors from a pool of experimentally synthesized compounds with known TP structures. 
However, this method cannot uncover TP framework compounds that remain unsynthesized. 
The recent synthesis of \ch{Li6SiO4Cl2} with a \ch{MgZn2}-type framework and \ch{Li6B7S13I} with a \ch{NaZn13}-type framework suggests a wealth of undiscovered TP framework compounds with potentially high ionic conductivity \citep{Morscher2021,Kaup2021}. 
To address this gap, we introduce a new rule-based scheme, the Met2Ion method, for generating the crystal structures of such unknown TP framework compounds, inspired by the structural relationship between metallic and ionic crystals.

\begin{figure}[t]%
\centering
\includegraphics[width=1\textwidth]{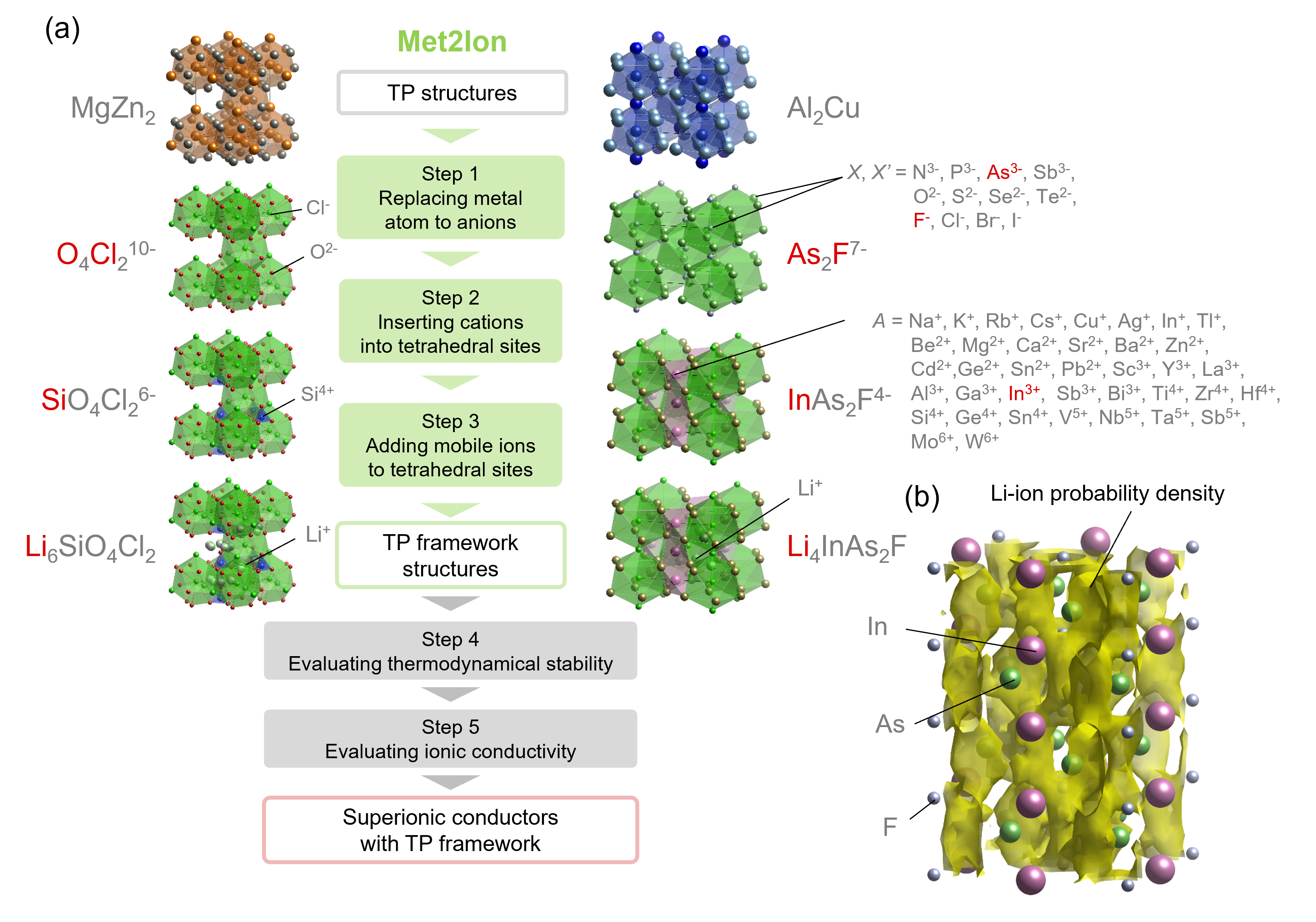}
\caption{(a) Search flow for tetrahedrally packed (TP) framework compounds using the Met2Ion method, which involves the following procedure to obtain \ch{Li6SiO4Cl2} from \ch{MgZn2} and the hypothetical \ch{Li4InAs2F} from \ch{Al2Cu}: Step 1 is to replace the metal atoms of the TP structures with two anions, $X$ and $X$', with candidate anions for $X$ and $X$' shown. In Step 2, a non-mobile cation $A$ is placed at the tetrahedral site, with the candidate cations for $A$ as shown. In Step 3, mobile cations are placed at the remaining tetrahedral sites to achieve electroneutrality. In Step 4, thermodynamically stable compounds are screened by performing ab initio calculations. In Step 5, compounds with high ionic conductivity are screened using first-principles molecular dynamics (FPMD) calculations. (b) Li-ion conductivity path of one of the compounds obtained, \ch{Li4InAs2F}.}\label{fig4}
\end{figure}

Our Met2Ion method employs a systematic process of elemental substitution and atomic arrangement, using metallic crystal structures as blueprints to create ionic crystal structures.
We applied this approach to discover novel, synthesizable Li-ion compounds within the TP framework that exhibit high ionic conductivity. 
The steps of the Met2Ion method, detailed in Figure \ref{fig4}a, include:

\begin{enumerate}
\item Replacing metal elements with anions in TP structures (e.g., \ch{MgZn2} to \ch{O4Cl2}$^{10-}$).
\item Inserting non-mobile cations into tetrahedral sites of the anionic framework (e.g., \ch{O4Cl2}$^{10-}$ to \ch{SiO4Cl2}$^{6-}$).
\item Adding mobile ions to tetrahedral sites to maintain electrical neutrality (e.g., \ch{SiO4Cl2}$^{6-}$ to \ch{Li6SiO4Cl2}).
\item Conducting structural optimization through first-principles calculations to assess thermodynamic stability.
\item Evaluating ionic conductivity using FPMD calculations.
\end{enumerate}

The rationale behind structure generation from Steps 1 to 3 is founded on the resemblance between TP metallic and TP ionic crystals. 
For instance, \ch{Li6SiO4Cl2} is derived from an \ch{MgZn2}-type framework by substituting metal atoms with Cl and O and introducing Li and Si at tetrahedral sites to achieve charge balance. 
This process is guided by the principle of substituting ions while preserving the elemental ratio seen in the metallic parent structure. 
Subsequently, we assess the thermodynamic stability and ionic conductivity of each candidate structure using first-principles calculations. 

Considering Figure \ref{fig3}a, which shows the \ch{Al2Cu}-type as a the second most common, our search extended to various Li$_xAX_2X^{\prime}$ compounds, with $A$ as a cation and $X$, $X^{\prime}$ as anions.
Table S3 and Figure S3 list the compounds identified through this screening. 
Six \ch{Al2Cu}-type framework compounds emerged as viable candidates for synthesis, with theoretical ionic conductivities surpassing 1.0$\times$10$^{-3}$ S/cm at room temperature and convex hull energies below 0.1 eV/atom. 
Notably, \ch{Li4InAs2F} exhibited a convex hull energy under 0.1 eV/atom and an impressive ionic conductivity of 8.0$\times$10$^{-2}$ S/cm. 
The crystal structure of \ch{Li4InAs2F}, depicted in Figure \ref{fig4}a, is characterized by an \ch{Al2Cu}-type framework forming a dodecahedron of As and F atoms. 
The primary Li-ion diffusion pathways in \ch{Li4InAs2F} run along the c-axis, as shown in Figure \ref{fig4}b, facilitated by the face-sharing connectivity of tetrahedral sites.

While we have focused on the \ch{Al2Cu}-type structure, our scheme is adaptable to other frameworks and mobile ions, holding promise for unveiling more high ionic conductors.
We recently established a method to generate crystal structures from polyhedra via discrete geometry analysis \cite{Yokoyama2023} which enables to generate new TP prototype structures.
The Met2Ion method surpasses traditional template-based prediction methods by enabling the discovery of ionic crystal structures that are not readily inferred from simple elemental substitutions \cite{Liu2023}. 
This study not only furnishes a blueprint for designing ionic conductive materials but also introduces an innovative method for predicting the structures of as-yet unknown TP framework compounds.

\section{Discussion}\label{sec3}
We have proposed a new design principle based on the fact that TP framework compounds exhibit high ionic conductivity, and through crystal structure design based on this principle, we have designed ionic conductors that could not be found via conventional compositional design methods.
First, we focused on the framework topology of Ag-ion conductors, which have been reported to have the highest available ionic conductivity.
We found that many Ag-ion conductors have a TP framework, along with a BCC framework, which was also known to allow high ionic conductivity.
Second, we showed that this design principle can be applied not only to Ag$^{+}$ ions, but also to other ions, including Li$^{+}$ and F$^-$. 
Therefore, TP framework compounds can serve as a universal structural design principle for the development of superionic conductors. 
Third, we have proposed a method for virtual screening of TP structures and TP framework compounds to aid in the discovery of new TP framework compounds. 
We have demonstrated that it is possible to discover new Li-ion conductors theoretically that are synthesizable and that show high ionic conductivity using our Met2Ion method. 
The results of this work offer the potential to create novel solid electrolytes with superionic conductivity for any mobile ionic species and to aid in the development of energy storage devices, including batteries and fuel cells.

\section{Metheds}\label{sec4}
\subsection*{Data preparation and analysis}\label{sec4}
The raw data of the crystal structures were exported from the Inorganic Crystal Structure Database (ICSD), version 2023.1, using the API \citep{Bergerhoff1983}. Among the structures that were labeled ``structure type'' in the database, structures composed of metal or metalloid elements and containing more than five data were extracted as ``ICSD prototype structures". The total number of prototype structures extracted was 621. Among the Ag-ion compounds in the ICSD, those with titles that included the words ``superionic", ``solid electrolyte", ``ion conductor", and ``ion conductivity" in the titles of their original papers were extracted, and data other than solid electrolyte materials were removed. Compounds with a framework that matched the prototype structure were then extracted as Ag-ion conductors using the StructureMatcher function \citep{Hundt2006} implemented in Pymatgen \citep{Ong2013}. Based on the charge information for the atoms in the structure file, negatively charged ions (anions) were considered as the framework ions, and positively charged ions (cations) were removed. We use StructureMatcher with $\texttt{ltol = 0.2}$, $\texttt{stol = 0.3}$, and $\texttt{angle\_tol = 2}$. If there are partially occupied sites in the structure, the sites were merged using the merge\_sites function with $\texttt{mod = average}$ and $\texttt{tol = 1}$. The total number of extracted Ag-ion conductors was 45. Ionic compounds with TP framework were extracted from ICSD under the same conditions. The visualizations were performed using VESTA, 3dt, and Systre software \citep{Momma2008,Delgado-Friedrichs2003}.

\subsection*{Ionic conductivity and probability density calculations}\label{sec4}
Density functional theory (DFT) calculations were performed using the Vienna Ab initio Simulation Package (VASP) with the projector augmented-wave approach and Perdew-Burke-Ernzerhof (PBE) generalized gradient approximation (GGA) functions \citep{Blochl1994, Perdew1996, Kresse1996, Kresse1999}. For the first-principles molecular dynamics (FPMD), a supercell model was created such that one side was larger than 8 \AA. Non-spin-polarized DFT calculations were performed using a k-point at the center of $\Gamma$. The time step was set to 2 fs. When the partially occupied structure was calculated, the method of Bushlanov \textit{et al.} \citep{Bushlanov2019} was used to obtain the most symmetrical atomic configuration, and this configuration was used as the initial structure. For the virtual structure, the lattice constants were fixed at those obtained after structural relaxation, but for the ICSD data, the lattice constants were fixed at the experimental values. The reason for the use of these constants is that the symmetry and the lattice constants after structural relaxation change depending on the initial atomic configuration when the partially occupied structure is calculated, which then affects the ion conductivity. Ion diffusion was evaluated using an NVT ensemble controlled by a Nosé-Hoover thermostat over a temperature range of 400 K to 900 K. The total time to convergence for the diffusivity ranged from 90 ps to 200 ps. Results at temperatures at which the framework structure melted ($\texttt{max\_framework\_displacement}$ was greater than 5 \AA) were removed. The first 5000 steps were removed to ensure thermal equilibrium, and the diffusion coefficients and ionic conductivities were calculated from the resulting mean-square displacement (MSD). The ion probability densities were obtained from FPMD calculations performed at the highest temperatures at which the framework did not collapse. The activation energies were obtained using the Nernst-Einstein equation. The ionic conductivity values were plotted versus the reciprocal of temperature to obtain the room-temperature ionic conductivity. The ionic probability density values within a structure were calculated by subdividing the supercell into a grid of cubic cells with an edge length of 0.2 \AA, and counting the number of time steps for which an ion occupies each cell. We used the diffusion analyzer in Pymatgen \citep{Ong2013} to perform these analyses.

\subsection*{Crystal structure generation and screening}\label{sec4}
The structure of an unknown TP framework compound was predicted using our proposed Met2Ion method, inspired by the structural rules for metallic and ionic crystals. Structure generation was performed in Steps 1, 2, and 3. In Step 1, the composition of the TP metallic crystal was maintained and the metallic atoms were replaced with anions. The Al and Cu atoms in \ch{Al2Cu} were replaced with two different anions. The candidate anions were N$^{3-}$, P$^{3-}$, As$^{3-}$, Sb$^{3-}$, O$^{2-}$, S$^{2-}$, Se$^{2-}$, Te$^{2-}$, F$^{-}$, Cl$^{-}$, Br$^{-}$ and I$^{-}$. In Step 2, the non-mobile cation was placed at the tetrahedral site with the lowest multiplicity of the Wyckoff label (site 4a in the \ch{Al2Cu} case). This is because the higher the number of non-mobile cations in the unit cell, the lower the ionic conductivity. The candidates for the non-mobile cations are: Na$^{+}$, K$^{+}$, Rb$^{+}$, Cs$^{+}$, Cu$^{+}$, Ag$^{+}$, In$^{+}$, Tl$^{+}$, Be$^{2+}$, Mg$^{2+}$, Ca$^{2+}$, Sr$^{2+}$, Ba$^{2+}$, Zn$^{2+}$, Cd$^{2+}$, Ge$^{2+}$, Sn$^{2+}$, Pb$^{2+}$, Sc$^{3+}$, Y$^{3+}$, La$^{3+}$, Al$^{3+}$, Ga$^{3+}$, In$^{3+}$, Sb$^{3+}$, Bi$^{3+}$, Ti$^{4+}$, Zr$^{4+}$, Hf$^{4+}$, Si$^{4+}$, Ge$^{4+}$, Sn$^{4+}$, V$^{5+}$, Nb$^{5+}$, Ta$^{5+}$, Sb$^{5+}$, Mo$^{6+}$, and W$^{6+}$. In Step 3, the movable ions were arranged to maintain electrical neutrality. Ideally, the most energetically stable atomic configuration should be identified from all combinations of atomic configurations. However, in view of the computational costs, the atomic configuration for the Li$^{+}$ ions in this work was determined from the most symmetrical structure using the method of Bushlanov \textit{et al.} \citep{Bushlanov2019}. Screening of the generated structures is performed in Steps 4 and 5. In Step 4, the thermodynamic stability of the crystal structure generated in Step 3 was evaluated via first-principles calculations. From a potential pool of 38 cations and 12 anions, a total of 5472 candidate compositions were calculated. The generated structures were subjected to structural relaxation and single-point calculations while symmetry was maintained. Competing phases were extracted from compounds with a convex hull energy of 0 eV/atom in the Materials Project \citep{Jain2013}, and structural relaxation and one-point calculations were also performed. DFT calculations were performed under essentially the same conditions that were used in the Materials Project \citep{Jain2013}. Structures with a convex hull energy of 0.1 eV/atom or less were selected as the synthesizable compounds \citep{Sun2016}. The number of candidate compounds extracted in Step 4 was 84. In Step 5, ionic conductivity was evaluated for the structures obtained in Step 4 by performing FPMD calculations. There were 6 candidate compounds obtained in Step 5 that theoretically had an ionic conductivity greater than 1.0$\times$10$^{-3}$ S/cm at room temperature. 

\backmatter

\bmhead{Supplementary information}

List of Ag-ion conductors extracted from ICSD; list of the calculated ionic conductivities and activation energies of the Ag-, Cu-, Li-, Na-, Zn-, Cd-, Mg-, and F-ion TP framework compounds; List of Li-ion conductor candidates with \ch{Al2Cu}-type framework screened using the Me2Ion method; Crystal structures of \ch{MgCu2} and \ch{Li8N2Te} with \ch{MgCu2}-type framework; Li-ion diffusion paths and mean square displacement (MSD) of Li ions of \ch{Li8N2Te} and \ch{Li7N2Te}; Arrhenius plots of \ch{Li7N2Te} obtained by FPMD calculations; and Arrhenius plots for virtual TP framework compounds obtained by FPMD calculations.

%\bmhead{Acknowledgments}

%Acknowledgments are not compulsory. Where included they should be brief. Grant or contribution numbers may be acknowledged.

%Please refer to Journal-level guidance for any specific requirements.

\bibliography{ref}% common bib file

\end{document}

\title[Article Title]{Supplementary Information:

Designing Superionic Conductors Using Tetrahedrally Packed Structures}


\author*[1]{\fnm{Tomoyasu} \sur{Yokoyama}}\email{yokoyama.tomoyasu@jp.panasonic.com}
\author[1]{\fnm{Kazuhide} \sur{Ichikawa}}
\author[1]{\fnm{Takuya} \sur{Naruse}}
\author[1]{\fnm{Kosei} \sur{Ohura}}
\author[1]{\fnm{Yukihiro} \sur{Kaneko}}

\affil*[1]{Technology Division, Panasonic Holdings Corporation,  1006 Kadoma, Kadoma City, Osaka 571-8508, Japan}

\maketitle
\clearpage

\begin{longtable}{ccccccc}
%\begin{minipage}{1000pt}
\caption{List of 45 Ag-ion conductors extracted from ICSD. ``ICSD ID" refers to the Collection Code of the database. ``Framework type" refers to the structural type of the anionic framework of each compound. }
\label{tab3}%
\\ \hline
ICSD ID & Formula &  Framework type \\
\hline
413508 & Ag$_{5}$Te$_{2}$Cl & \ch{Al2Cu} \\
83282 & Ag$_{8}$TiS$_{6}$ & \ch{MgCu2}  \\
54056 & Ag$_{7}$PSe$_{6}$ & \ch{MgCu2}  \\
241154 & Ag$_{7}$GeSe$_{5}$I & \ch{MgCu2}  \\
100267 & Ag$_{8}$GeTe$_{6}$ & \ch{MgCu2}  \\
51487 & Ag$_{6.684}$GeSe$_{5}$I$_{0.69}$ & \ch{MgCu2}  \\
51488 & Ag$_{6.72}$GeSe$_{5}$I$_{0.69}$ & \ch{MgCu2}  \\
40957 & Ag$_{7}$NbS$_{6}$ & \ch{MgCu2}  \\
29704 & Cu$_{1.5}$Ag$_{1.5}$PS$_{4}$ & \ch{Li2BaGe}  \\
29706 & Ag$_{3}$PS$_{4}$ & \ch{Li2BaGe}  \\
173113 & Ag$_{10}$Te$_{4}$Br$_{3}$ & \ch{Zr4Al3}  \\
159854 & Ag$_{3}$SI$_{0.4}$Br$_{0.6}$ & \ch{BCC}  \\
93448 & AgHg$_{0.5}$I$_{2}$ & \ch{BCC}  \\
159851 & Ag$_{3}$SI$_{0.8}$Br$_{0.2}$ & \ch{BCC}  \\
159852 & Ag$_{3}$SI$_{0.67}$Br$_{0.33}$ & \ch{BCC}  \\
159857 & Ag$_{3}$SI$_{0.05}$Br$_{0.95}$ & \ch{BCC}  \\
159855 & Ag$_{3}$SI$_{0.25}$Br$_{0.75}$ & \ch{BCC}  \\
159856 & Ag$_{3}$SI$_{0.17}$Br$_{0.83}$ & \ch{BCC}  \\
159853 & Ag$_{3}$SI$_{0.6}$Br$_{0.4}$ & \ch{BCC}  \\
93429 & Ag$_{3}$SI & \ch{BCC}  \\
2108 & AgI & \ch{BCC}  \\
93428 & Ag$_{3}$SI & \ch{BCC}  \\
200256 & AgI & \ch{BCC}  \\
201003 & Ag$_{2.88}$SI & \ch{BCC}  \\
9587 & Ag$_{2}$S & \ch{BCC}  \\
9586 & Ag$_{2}$S & \ch{BCC}  \\
93430 & Ag$_{3}$SI & \ch{BCC}  \\
174096 & Ag$_{3}$SBr & \ch{BCC}  \\
190397 & Ag$_{2}$S & \ch{FCC}  \\
95608 & Ag$_{0.667}$Pb$_{0.167}$I & \ch{FCC}  \\
95609 & Ag$_{0.5}$Pb$_{0.25}$I & \ch{FCC}  \\
190590 & Ag$_{2}$$_{.}$$_{4}$Sn$_{0}$$_{.}$$_{8}$I$_{4}$ & \ch{FCC}  \\
190587 & Ag$_{2}$CdI$_{4}$ & \ch{FCC}  \\
190588 & Ag$_{2}$ZnI$_{4}$ & \ch{HCP}  \\
190589 & AgZn$_{0.5}$I$_{2}$ & \ch{HCP}  \\
93447 & AgHg$_{0.5}$I$_{2}$ & \ch{HCP}  \\
601918 & Ag$_{0.144}$Ga$_{1.286}$S$_{2}$ & \ch{HCP}  \\
51909 & RbAg$_{4}$I$_{5}$ & $\beta$-\ch{Mn}  \\
41591 & RbAg$_{4}$I$_{5}$ & $\beta$-\ch{Mn}  \\
33403 & CsAg$_{3.482}$I$_{2.999}$Br$_{1.998}$ & $\beta$-\ch{Mn}  \\
33401 & CsAg$_{3.668}$I$_{2.33}$Br$_{2.67}$ & $\beta$-\ch{Mn}  \\
33402 & CsAg$_{3.598}$I$_{2.669}$Br$_{2.331}$ & $\beta$-\ch{Mn}  \\
33400 & CsAg$_{3.446}$I$_{2.237}$Br$_{2.763}$ & $\beta$-\ch{Mn}  \\
27203 & RbAg$_{4}$I$_{5}$ & $\beta$-\ch{Mn}  \\
51910 & KAg$_{3.925}$I$_{5}$ & $\beta$-\ch{Mn}  \\
\hline
\end{longtable}

\newpage

\begin{figure}[htbp]%
\centering
\includegraphics[width=1\textwidth]{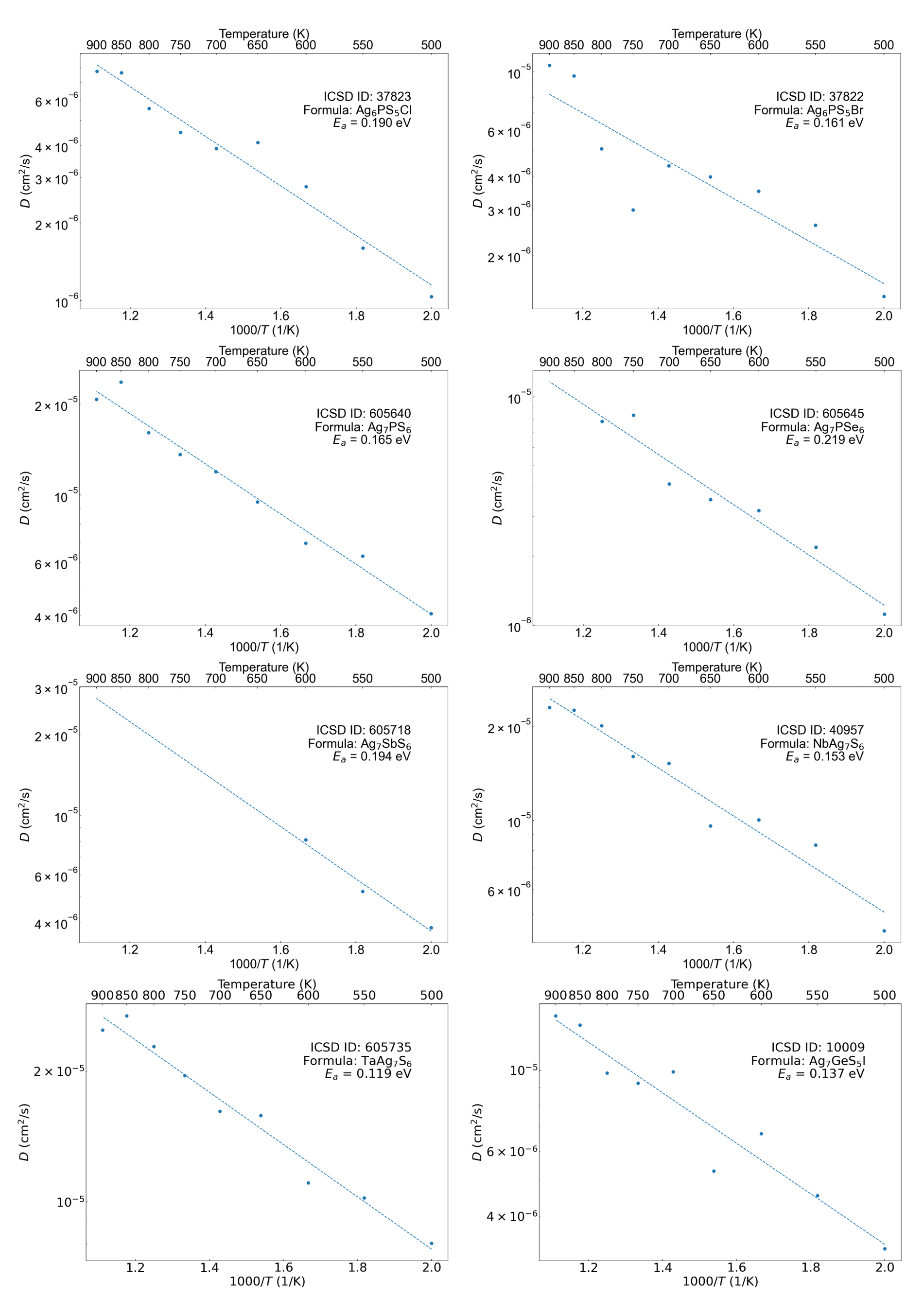}
\end{figure}
\begin{figure}[htbp]%
\centering
\includegraphics[width=1\textwidth]{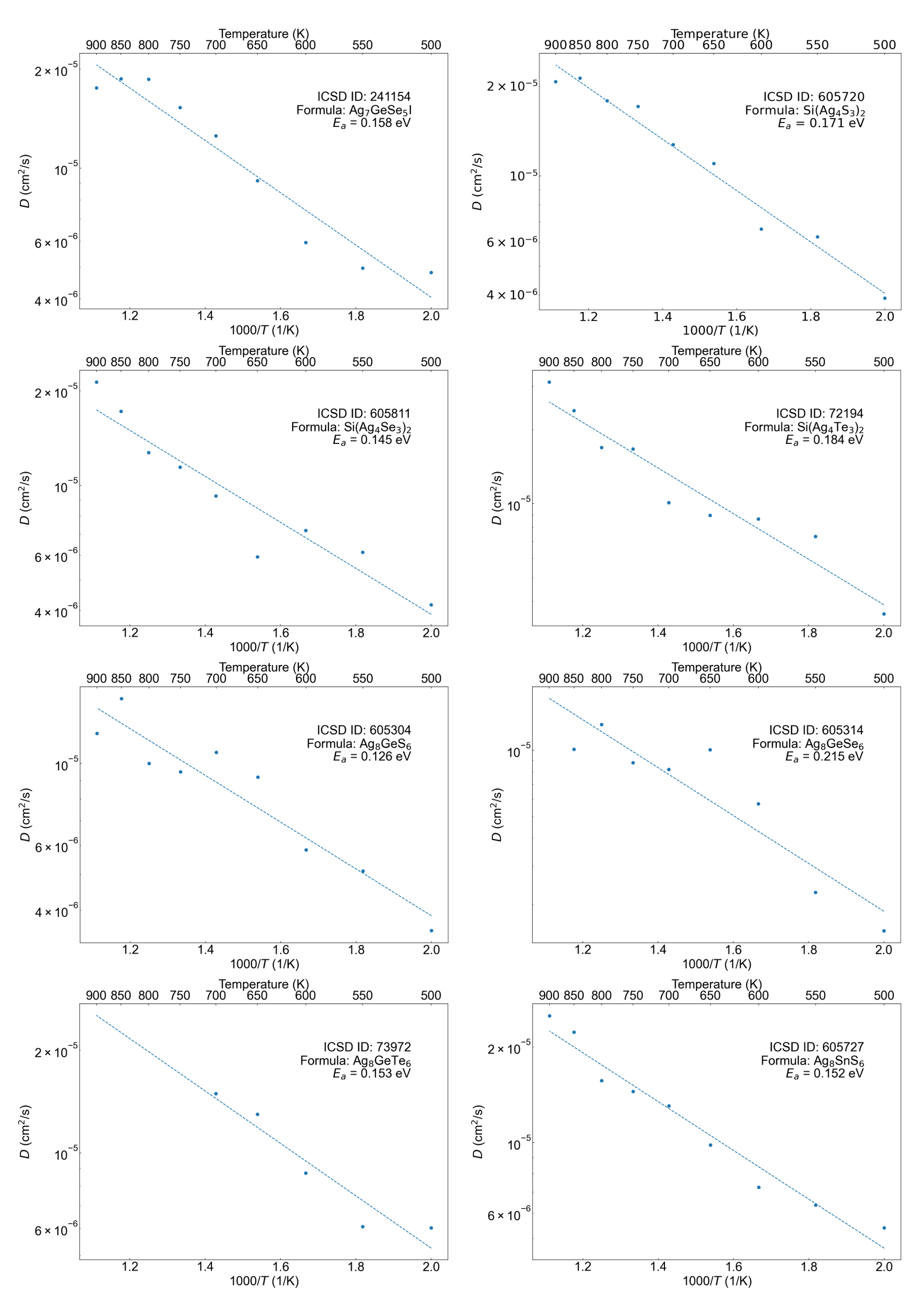}
\end{figure}
\begin{figure}[htbp]%
\centering
\includegraphics[width=1\textwidth]{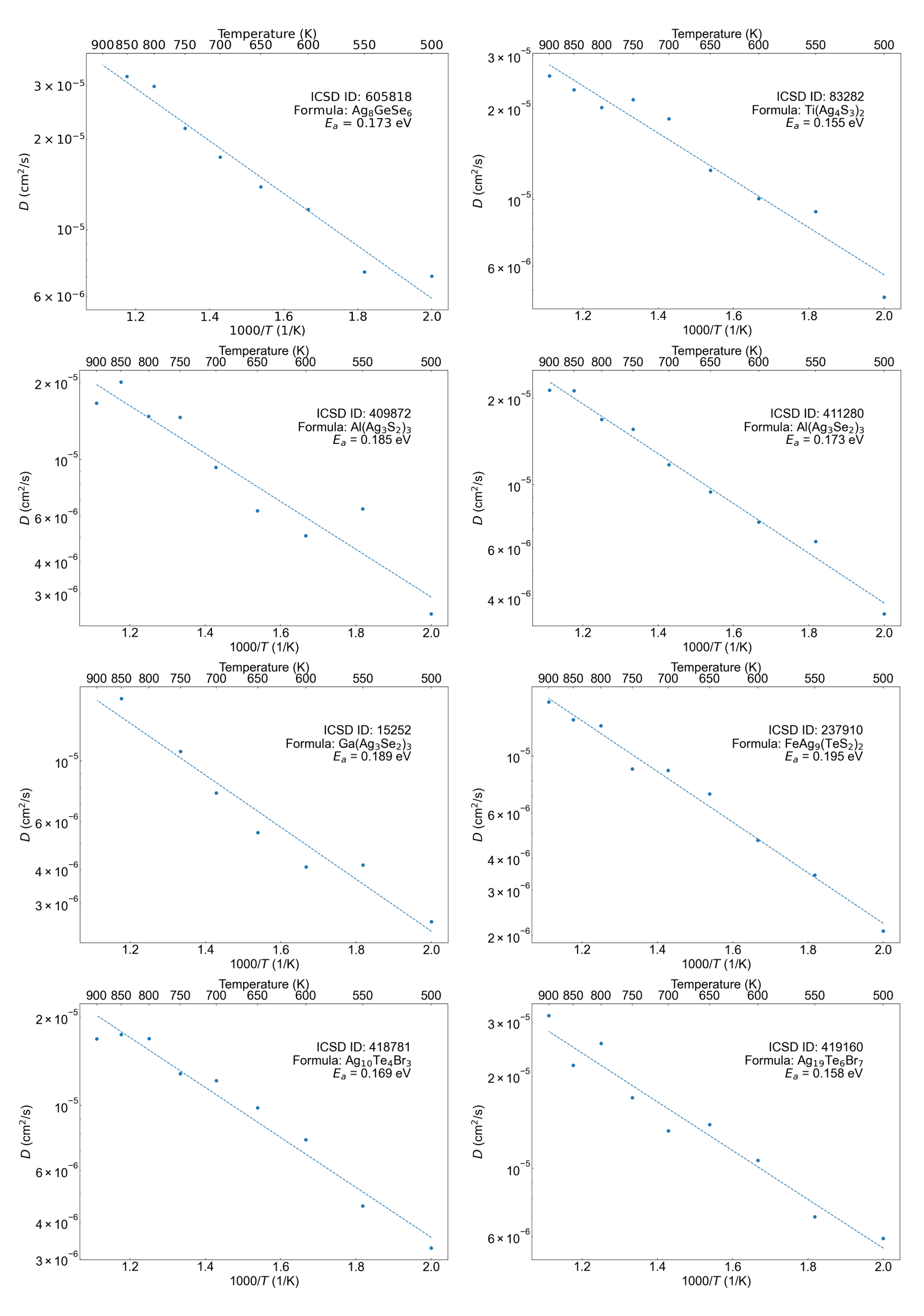}
\end{figure}
\begin{figure}[htbp]%
\centering
\includegraphics[width=1\textwidth]{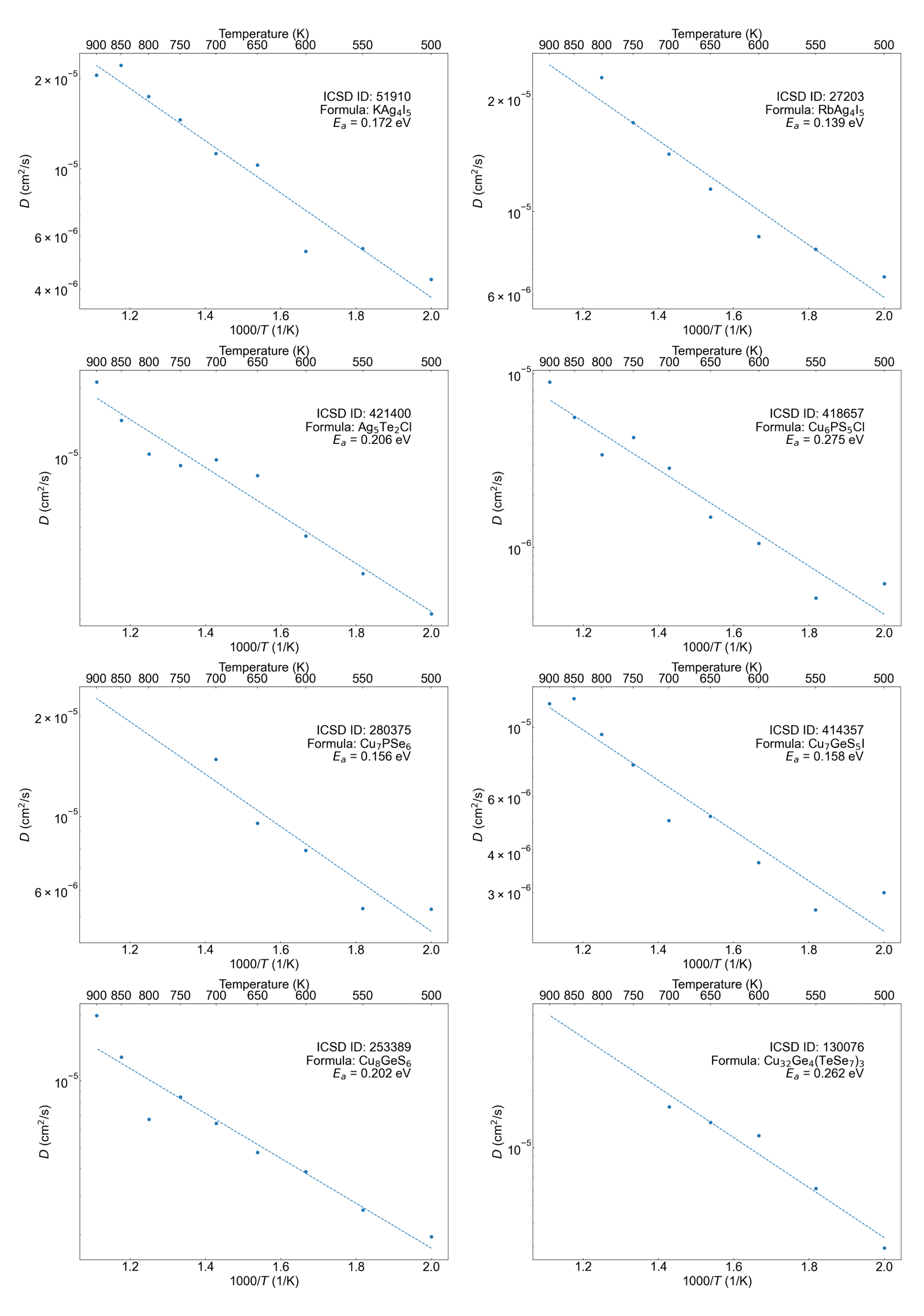}
\end{figure}
\begin{figure}[htbp]%
\centering
\includegraphics[width=1\textwidth]{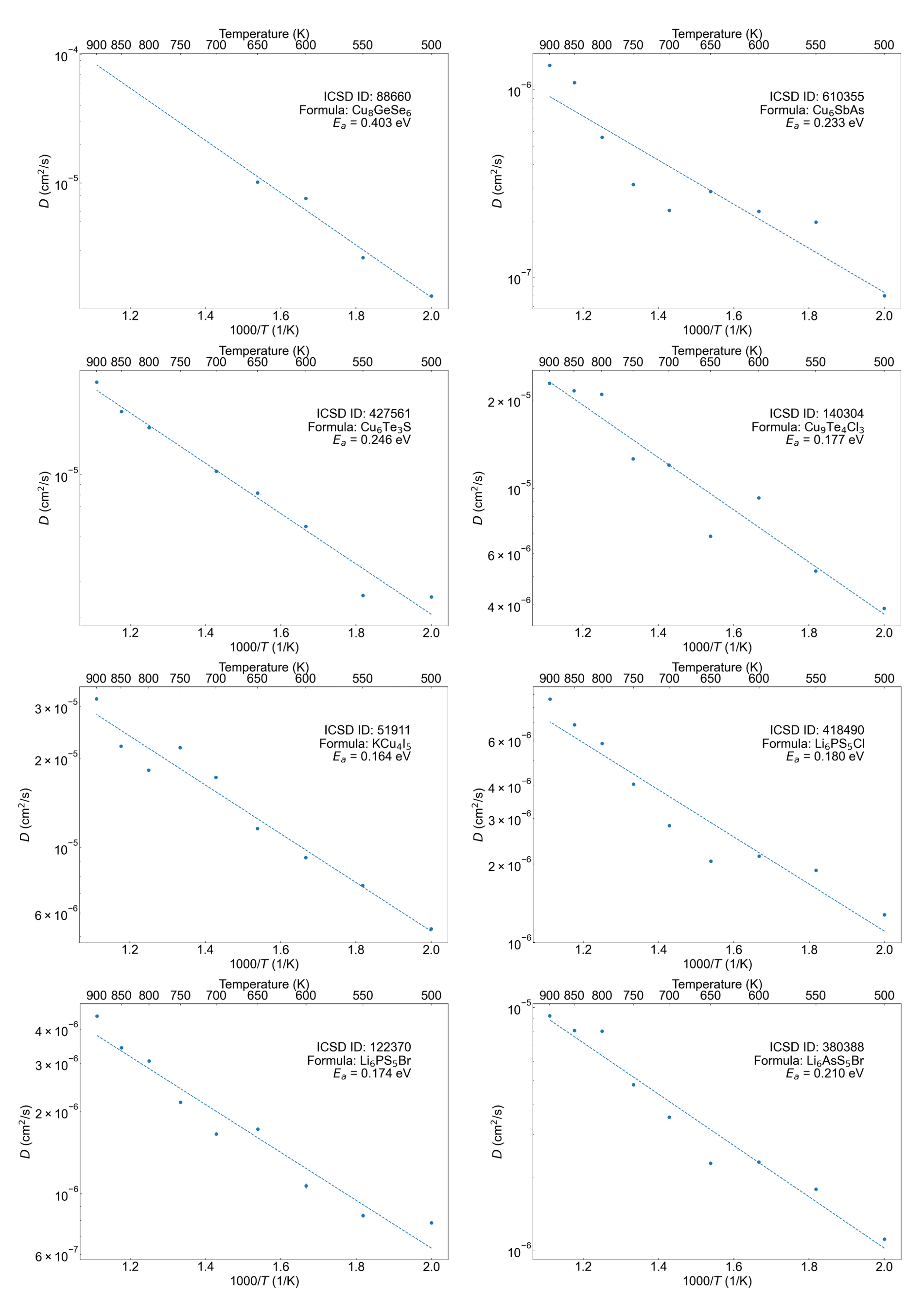}
\end{figure}
\begin{figure}[htbp]%
\centering
\includegraphics[width=1\textwidth]{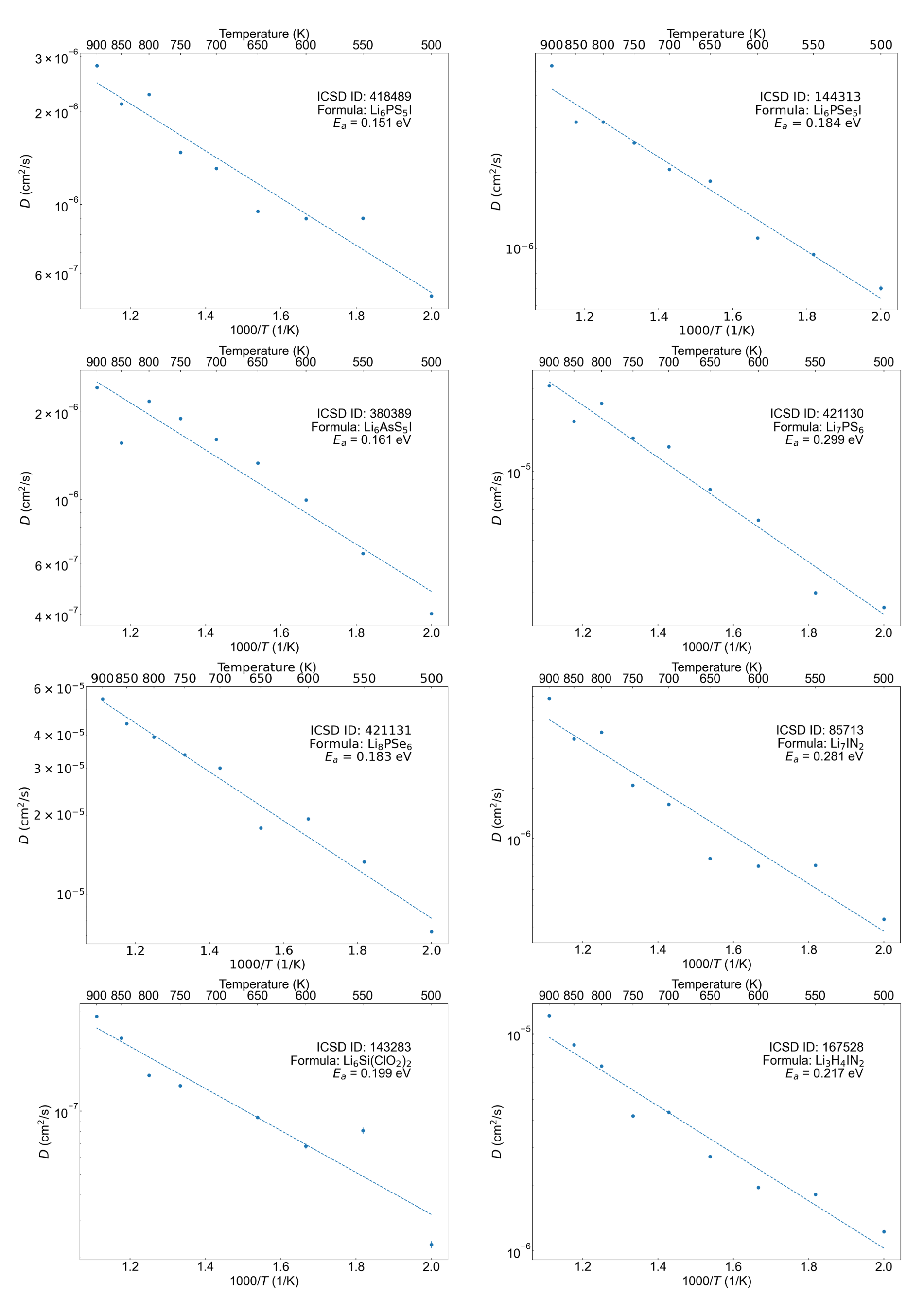}
\end{figure}
\begin{figure}[htbp]%
\centering
\includegraphics[width=1\textwidth]{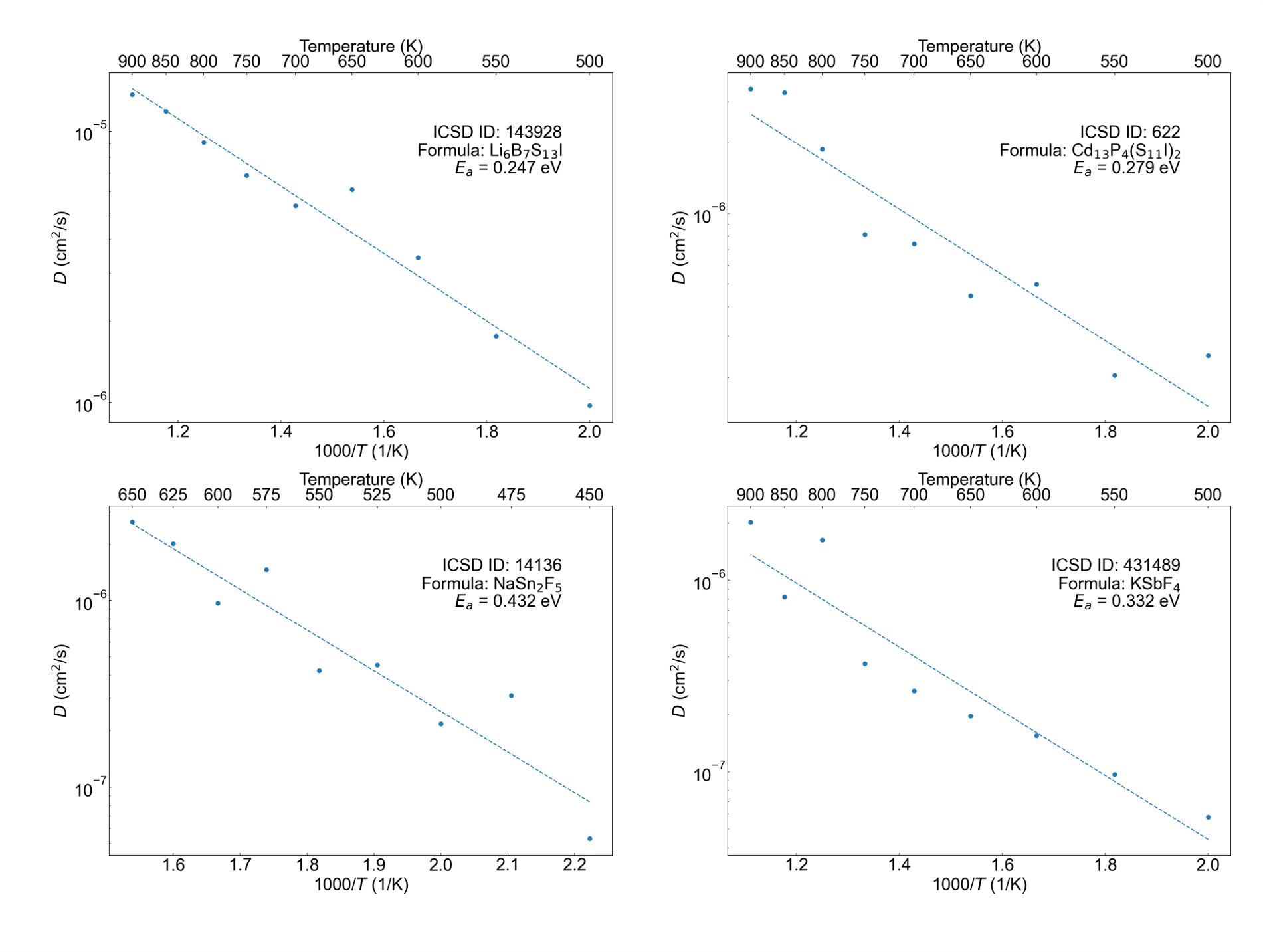}
\caption{Arrhenius plots for TP framework compounds in the ICSD obtained by first-principles molecular dynamics calculations.}\label{figS1}
\end{figure}

\newpage

\begin{longtable}{cccccc}
%\begin{minipage}{1000pt}
\caption{List of the calculated ionic conductivities and activation energies of the Ag-, Cu-, Li-, Na-, Zn-, Cd-, Mg-, and F-ion TP framework compounds. ``Ion" refers to a possible mobile ion in the compound. ``Formula" refers to the compositional formula of the calculated structure. ``ICSD ID" refers to the database collection code. ``Framework type" refers to the structural type of the anion framework for each compound. $\sigma_{300K}$ and $E_{a,300K}$ mean the ionic conductivity at room temperature and the activation energy as extrapolated from Arrhenius plots based on FPMD calculations, respectively. Estimates of the upper and lower limits of the ionic conductivity and the activation energy are included. If the ion conductivity is difficult to evaluate because the ions hardly diffuse, the ion conductivity and activation energy columns are labeled ``Too Low" and ``Too High", respectively.}
\label{tab3}%
\\ \hline
Ion & ICSD ID & Formula & Framework type & $\sigma_{300K}$ (mS/cm) & $E_{a,300K}$ (eV) \\
\hline
{Ag}$^{+}$ & 37823 & \ch{Ag6PS5Cl} & \ch{MgCu2} & 8.1 (4.8, 13.8)&0.190$\pm$0.014\\
{Ag}$^{+}$ & 37822 & \ch{Ag6PS5Br} & \ch{MgCu2} & 16.3 (5.0, 53.2)&0.161$\pm$0.030\\
{Ag}$^{+}$ & 605640 & \ch{Ag7PS6} & \ch{MgCu2} & 47.4 (32.4, 69.4)&0.165$\pm$0.010\\
{Ag}$^{+}$ & 605645 & \ch{Ag7PSe6} & \ch{MgCu2} & 5.8 (2.4, 13.9)&0.219$\pm$0.023\\
{Ag}$^{+}$ & 605718 & \ch{Ag7SbS6} & \ch{MgCu2} & 26.3 (8.4, 82.9)&0.194$\pm$0.030\\
{Ag}$^{+}$ & 40957 & \ch{Ag7NbS6} & \ch{MgCu2} & 71.1 (43.9, 115.0)&0.153$\pm$0.012\\
{Ag}$^{+}$ & 605735 & \ch{Ag7TaS6} & \ch{MgCu2} & 183.4 (138.0, 243.8)&0.119$\pm$0.007\\
{Ag}$^{+}$ & 10009 & \ch{Ag7GeS5I} & \ch{MgCu2} & 56.8 (31.8, 101.6)&0.137$\pm$0.015\\
{Ag}$^{+}$ & 241154 & \ch{Ag7GeSe5I} & \ch{MgCu2} & 46.3 (25.3, 84.8)&0.158$\pm$0.016\\
{Ag}$^{+}$ & 605720 & \ch{Ag8SiS6} & \ch{MgCu2} & 47.6 (31.9, 71.1)&0.171$\pm$0.010\\
{Ag}$^{+}$ & 605811 & \ch{Ag8SiSe6} & \ch{MgCu2} & 61.7 (30.6, 124.5)&0.145$\pm$0.018\\
{Ag}$^{+}$ & 72194 & \ch{Ag8SiTe6} & \ch{MgCu2} & 29.0 (14.3, 58.8)&0.184$\pm$0.018\\
{Ag}$^{+}$ & 605304 & \ch{Ag8GeS6} & \ch{MgCu2} & 89.0 (49.2, 161.1)&0.126$\pm$0.015\\
{Ag}$^{+}$ & 605314 & \ch{Ag8GeSe6} & \ch{MgCu2} & 10.1 (2.4, 41.7)&0.215$\pm$0.037\\
{Ag}$^{+}$ & 73972 & \ch{Ag8GeTe6} & \ch{MgCu2} & 63.5 (21.3, 189.1)&0.153$\pm$0.028\\
{Ag}$^{+}$ & 605727 & \ch{Ag8SnS6} & \ch{MgCu2} & 68.5 (44.3, 105.9)&0.152$\pm$0.011\\
{Ag}$^{+}$ & 605818 & \ch{Ag8SnSe6} & \ch{MgCu2} & 59.2 (35.5, 98.8)&0.173$\pm$0.013\\
{Ag}$^{+}$ & 83282 & \ch{Ag8TiS6} & \ch{MgCu2} & 83.8 (50.6, 138.8)&0.155$\pm$0.013\\
{Ag}$^{+}$ & 409872 & \ch{Ag9AlS6} & \ch{MgCu2} & 30.1 (12.6, 72.0)&0.185$\pm$0.023\\
{Ag}$^{+}$ & 411280 & \ch{Ag9AlSe6} & \ch{MgCu2} & 43.4 (32.3, 58.3)&0.173$\pm$0.008\\
{Ag}$^{+}$ & 15252 & \ch{Ag9GaSe6} & \ch{MgCu2} & 20.6 (9.4, 45.2)&0.189$\pm$0.020\\
{Ag}$^{+}$ & 237910 & \ch{Ag9FeTe2S4} & \ch{MgCu2} & 18.0 (13.1, 24.9)&0.195$\pm$0.008\\
{Ag}$^{+}$ & 418781 & \ch{Ag10Te4Br3} & \ch{Zr4Al3} & 38.2 (24.0, 61.0)&0.169$\pm$0.012\\
{Ag}$^{+}$ & 419160 & \ch{Ag19Te6Br7} & \ch{W6Fe7} & 71.6 (43.1, 118.9)&0.158$\pm$0.013\\
{Ag}$^{+}$ & 51910 & \ch{KAg4I5} & $\beta$-Mn & 18.6 (10.4, 33.1)&0.172$\pm$0.015\\
{Ag}$^{+}$ & 27203 & \ch{RbAg4I5} & $\beta$-Mn & 48.3 (26.5, 88.0)&0.139$\pm$0.016\\
{Ag}$^{+}$ & 421400 & \ch{Ag5Te2Cl} & \ch{Al2Cu} & 14.8 (7.7, 28.1)&0.206$\pm$0.017\\
{Ag}$^{+}$ & 18540 & \ch{K2Ag2SnS4} & \ch{CrB} & Too low&Too high\\
{Ag}$^{+}$ & 80195 & \ch{Cs2Ag2ZrTe4} & \ch{CrB} & Too low&Too high\\
{Ag}$^{+}$ & 55840 & \ch{CsAgZnS2} & \ch{CrB} & Too low&Too high\\
{Cu}$^{+}$ & 418657 & \ch{Cu6PS5Cl} & \ch{MgCu2} & 1.0 (0.3, 2.9)&0.275$\pm$0.028\\
{Cu}$^{+}$ & 33503 & \ch{Cu6PS5Br} & \ch{MgCu2} & Too low&Too high\\
{Cu}$^{+}$ & 415489 & \ch{Cu6P1S5I} & \ch{MgCu2} & Too low&Too high\\
{Cu}$^{+}$ & 280375 & \ch{Cu7PSe6} & \ch{MgCu2} & 67.6 (19.3, 236.5)&0.156$\pm$0.032\\
{Cu}$^{+}$ & 414357 & \ch{Cu7GeS5I} & \ch{MgCu2} & 33.9 (16.9, 67.7)&0.158$\pm$0.018\\
{Cu}$^{+}$ & 253389 & \ch{Cu8GeS6} & \ch{MgCu2} & 15.2 (6.6, 35.0)&0.202$\pm$0.022\\
{Cu}$^{+}$ & 253394 & \ch{Cu8GeSe6} & \ch{MgCu2} & Too low&Too high\\
{Cu}$^{+}$ & 130076 & \ch{Cu8GeTe2Se4} & \ch{MgCu2} & 9.9 (2.7, 35.9)&0.262$\pm$0.033\\
{Cu}$^{+}$ & 88660 & \ch{Cu8GeSe6} & \ch{MgZn2} & 0.5 (0.1, 3.0)&0.403$\pm$0.048\\
{Cu}$^{+}$ & 610355 & \ch{Cu6SbAs} & \ch{Cr3Si} & 0.8 (0.2, 3.2)&0.233$\pm$0.037\\
{Cu}$^{+}$ & 427561 & \ch{Cu6Te3S} & \ch{Cr3Si} & 8.5 (4.6, 16.0)&0.246$\pm$0.016\\
{Cu}$^{+}$ & 140304 & \ch{Cu9Te4Cl3} & \ch{Zr4Al3} & 40.9 (20.1, 83.1)&0.177$\pm$0.018\\
{Cu}$^{+}$ & 51911 & \ch{KCu4I5} & $\beta$-Mn & 31.9 (19.6, 51.8)&0.164$\pm$0.013\\
{Cu}$^{+}$ & 201347 & \ch{Cu3B7O13Br} & \ch{NaZn13} & Too low&Too high\\
{Cu}$^{+}$ & 61058 & \ch{Cu3B7O13I} & \ch{NaZn13} & Too low&Too high\\
{Li}$^{+}$ & 421479 & \ch{Li6PO5Cl} & \ch{MgCu2} & Too low&Too high\\
{Li}$^{+}$ & 421480 & \ch{Li6PO5Br} & \ch{MgCu2} & Too low&Too high\\
{Li}$^{+}$ & 418490 & \ch{Li6PS5Cl} & \ch{MgCu2} & 10.6 (4.7, 23.9)&0.180$\pm$0.021\\
{Li}$^{+}$ & 122370 & \ch{Li6PS5Br} & \ch{MgCu2} & 6.6 (3.6, 11.9)&0.174$\pm$0.015\\
{Li}$^{+}$ & 380388 & \ch{Li6AsS5Br} & \ch{MgCu2} & 5.7 (2.9, 11.2)&0.210$\pm$0.017\\
{Li}$^{+}$ & 418489 & \ch{Li6PS5I} & \ch{MgCu2} & 7.1 (3.9, 12.9)&0.151$\pm$0.015\\
{Li}$^{+}$ & 144313 & \ch{Li6PSe5I} & \ch{MgCu2} & 4.7 (2.9, 7.8)&0.184$\pm$0.013\\
{Li}$^{+}$ & 380389 & \ch{Li6AsS5I} & \ch{MgCu2} & 5.5 (2.7, 11.5)&0.161$\pm$0.019\\
{Li}$^{+}$ & 421130 & \ch{Li7PS6} & \ch{MgCu2} & 2.5 (1.1, 5.9)&0.299$\pm$0.022\\
{Li}$^{+}$ & 421131 & \ch{Li7PSe6} & \ch{MgCu2} & 79.6 (49.8, 127.0)&0.183$\pm$0.012\\
{Li}$^{+}$ & 85713 & \ch{Li7N2I} & \ch{MgCu2} & 1.1 (0.4, 3.7)&0.281$\pm$0.030\\
{Li}$^{+}$ & 247255 & \ch{Li8N2Se} & \ch{MgCu2} & Too low&Too high\\
{Li}$^{+}$ & 247258 & \ch{Li8N2Te} & \ch{MgCu2} & Too low&Too high\\
{Li}$^{+}$ & 143283 & \ch{Li6SiO4Cl2} & \ch{MgZn2} & 0.3 (0.1, 1.0)&0.199$\pm$0.027\\
{Li}$^{+}$ & 167528 & \ch{Li3(NH2)2I} & \ch{MgZn2} & 5.3 (2.5, 11.3)&0.217$\pm$0.019\\
{Li}$^{+}$ & 402881 & \ch{LiGa3Te5} & $\beta$-Mn & Too low&Too high\\
{Li}$^{+}$ & 55064 & \ch{Li(NH3)I} & \ch{CrB} & Too low&Too high\\
{Li}$^{+}$ & 74930 & \ch{Li2(OH)I} & \ch{FeB} & Too low&Too high\\
{Li}$^{+}$ & 143928 & \ch{Li6B7S13I} & \ch{NaZn13} & 2.1 (1.1, 4.1)&0.247$\pm$0.017\\
{Na}$^{+}$ & 253640 & \ch{Na2CdGe2Se6} & \ch{MgZn2} & Too low&Too high\\
{Na}$^{+}$ & 402329 & \ch{NaGa3Te5} & $\beta$-Mn & Too low&Too high\\
{Na}$^{+}$ & 243562 & \ch{NaIn2GaSe5} & $\beta$-Mn & Too low&Too high\\
{Na}$^{+}$ & 402879 & \ch{NaGa6AgTe10} & $\beta$-Mn & Too low&Too high\\
{Na}$^{+}$ & 12964 & \ch{CsNaMnTe2} & \ch{CrB} & Too low&Too high\\
{Na}$^{+}$ & 47100 & \ch{NaGe2N3} & \ch{CaCu5} & Too low&Too high\\
{Zn}$^{2+}$ & 420783 & \ch{ZnHg3S2Cl4} & \ch{MgZn2} & Too low&Too high\\
{Zn}$^{2+}$ & 65257 & \ch{Cs2Mn2ZnS4} & \ch{CrB} & Too low&Too high\\
{Zn}$^{2+}$ & 55444 & \ch{Zn3B7O13Cl} & \ch{NaZn13} & Too low&Too high\\
{Cd}$^{2+}$ & 622 & \ch{Cd13P4S22I2} & \ch{MgCu2} & 0.7 (0.1, 2.9)&0.279$\pm$0.039\\
{Cd}$^{2+}$ & 84983 & \ch{Cd8As7Cl} & \ch{Cr3Si} & Too low&Too high\\
{Cd}$^{2+}$ & 85581 & \ch{CsCdAuS2} & \ch{CrB} & Too low&Too high\\
{Mg}$^{2+}$ & 22009 & \ch{Mg3B7O13Cl} & \ch{NaZn13} & Too low&Too high\\
{F}$^{-}$ & 14136 & \ch{NaSn2F5} & \ch{Al2Cu} & 1.8 (0.2, 14.9)&0.432$\pm$0.055\\
{F}$^{-}$ & 431489 & \ch{KSbF4} & \ch{FeB} & 1.5 (0.3, 8.1)&0.332$\pm$0.043\\
\hline
\end{longtable}

\newpage

\begin{longtable}{cccccc}
%\begin{minipage}{1000pt}
\caption{List of Li-ion conductor candidates screened using the Me2Ion method under the conditions that $E_{hull}$ $<$ 0.1 eV/atom and $\sigma_{300K}$ $>$ 1$\times$10$^{-3}$ S/cm. ``Ion" refers to a possible mobile ion in the compound. ``Formula" refers to the compositional formula of the calculated structure. ``Framework type" refers to the structural type of the anionic framework for each compound. $E_{hull}$ represents the convex hull energy, which is an index of the thermodynamic stability obtained from first-principles calculations. $\sigma_{300K}$ and $E_{a,300K}$ mean the ionic conductivity at room temperature and the activation energy as extrapolated from Arrhenius plots based on FPMD calculations, respectively.}\label{tab3}%
\\ \hline
Ion &  Formula & Framework type & $E_{hull}$  (eV/atom) & $\sigma_{300K}$ (S/cm) & $E_{a,300K}$ (eV)\\
\hline
Li$^{+}$ &  \ch{Li4InAs2F}	& \ch{Al2Cu} & 0.099 &8.0$\times10^{-2}$ & 0.171 \\
Li$^{+}$ &  \ch{Li4InP2F}	& \ch{Al2Cu} & 0.095 &3.6$\times10^{-2}$ & 0.204 \\
Li$^{+}$ &  \ch{Li4InSb2Cl}	& \ch{Al2Cu} & 0.091 &3.1$\times10^{-2}$ & 0.192 \\
Li$^{+}$ &  \ch{Li3SiP2Cl}	& \ch{Al2Cu} & 0.093 &1.2$\times10^{-2}$ & 0.242 \\
Li$^{+}$ & \ch{Li4InAs2Cl}	& \ch{Al2Cu} & 0.077 &2.0$\times10^{-3}$ & 0.294 \\
Li$^{+}$ &  \ch{Li4AlAs2F}	& \ch{Al2Cu} & 0.072 &1.6$\times10^{-3}$ & 0.330 \\
\hline
\end{longtable}

\newpage

\begin{figure}[h]%
\centering
\includegraphics[width=1\textwidth]{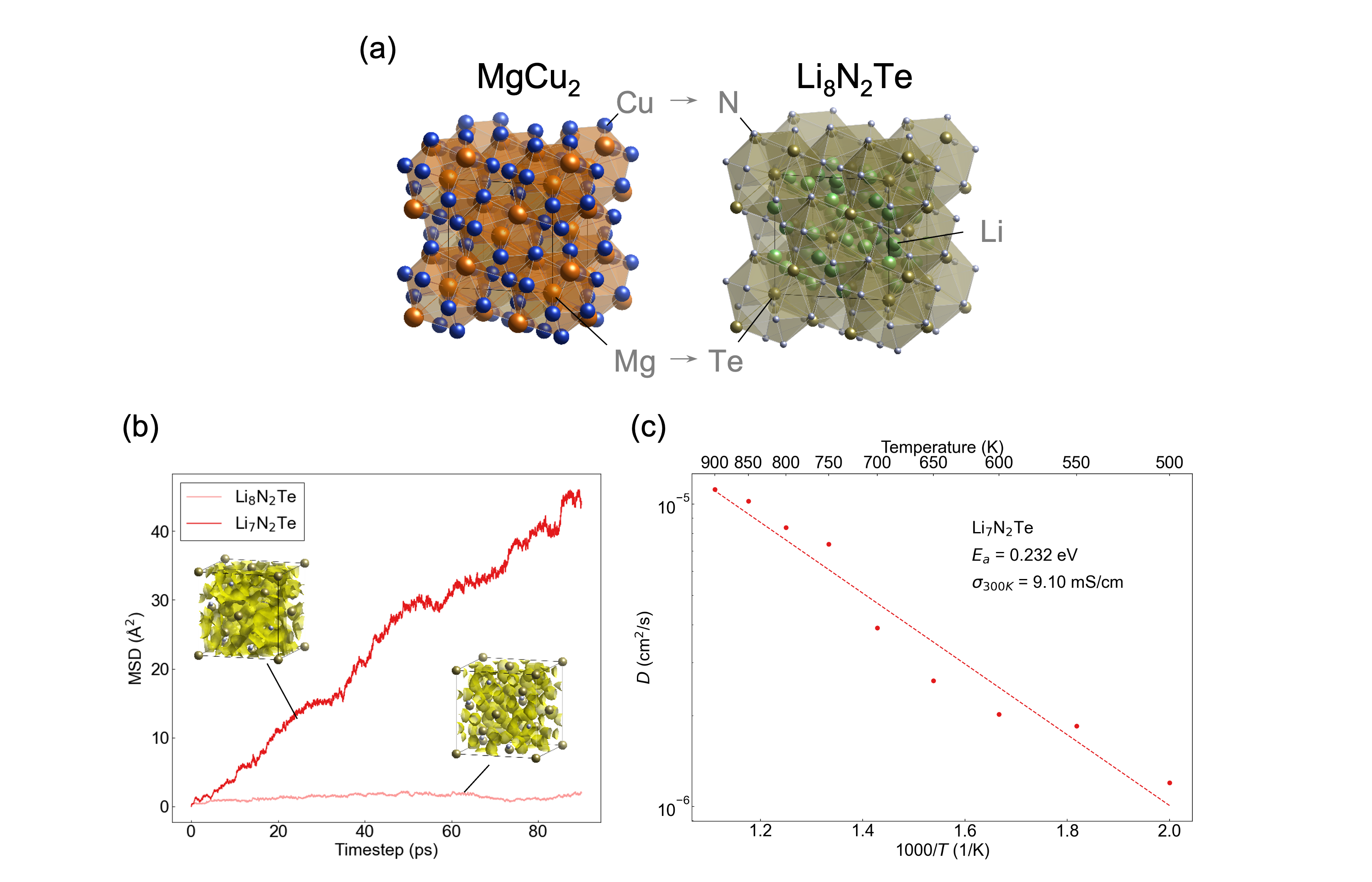}
\caption{(a) Crystal structures of \ch{MgCu2} and \ch{Li8N2Te} with \ch{MgCu2}-type framework. (b) Li-ion diffusion paths and mean square displacement (MSD) of Li ions of \ch{Li8N2Te} and \ch{Li7N2Te}, and (c) Arrhenius plots of \ch{Li7N2Te} obtained by first-principles molecular dynamics calculations.}\label{figS2}
\end{figure}

\newpage

\begin{figure}[h]%
\centering
\includegraphics[width=1\textwidth]{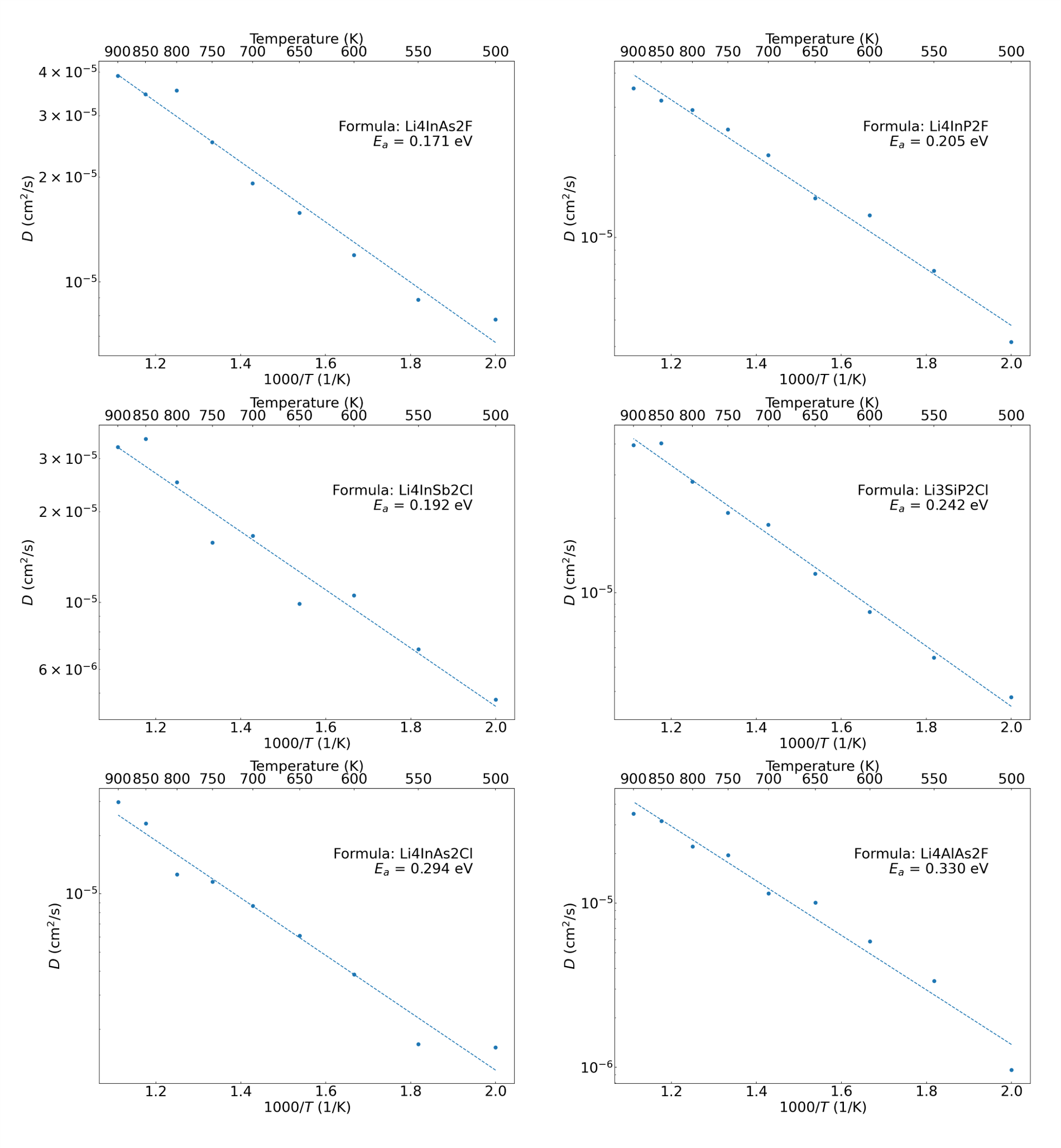}
\caption{Arrhenius plots for Li-ion conductor candidates with \ch{Al2Cu}-type framework listed in Table S3 obtained by first-principles molecular dynamics calculations.}\label{figS3}
\end{figure}

%\bibliography{ref}% common bib file